\documentclass[groupedaddress,nofootinbib,11pt,preprintnumbers]{revtex4-2}
\usepackage[utf8]{inputenc}
\usepackage[table,dvipsnames]{xcolor}
\usepackage[colorlinks,linktocpage]{hyperref}
\hypersetup{
	citecolor  = NavyBlue,
	linkcolor  = NavyBlue,
	urlcolor = NavyBlue
}

\usepackage{slashed}
\usepackage{graphicx}

\graphicspath{{./figs/}{../figs/}}

\usepackage{bbold}
\usepackage{amsmath}
\usepackage{amssymb}
\usepackage{latexsym}
\usepackage{mathtools}
\usepackage{dsfont}
\usepackage{hyperref}
\usepackage{amsfonts}
\usepackage{enumitem}
\usepackage{multirow}
\usepackage{xstring} 
\usepackage{xspace} 
\usepackage[normalem]{ulem}
\usepackage{fontawesome}
\usepackage{cancel}
\usepackage{bm}
\usepackage[most]{tcolorbox}
\usepackage{pifont}
\newcommand{\xmark}{\ding{55}}

\def\vect#1{\boldsymbol{#1}}

\def\ie{{\it i.e.}}
\def\eg{{\it e.g.}}

\newcommand\Tstrut{\rule{0pt}{3.5ex}}         % = `top' strut
\newcommand\Bstrut{\rule[-2ex]{0pt}{0pt}}   % = `bottom' strut

\usepackage{feynmp-auto}
\begin{document}

\title{Dark Matter Absorption via Electronic Excitations}

\author{Andrea Mitridate, Tanner Trickle, Zhengkang Zhang, Kathryn M. Zurek}
\affiliation{Walter Burke Institute for Theoretical Physics, California Institute of Technology, Pasadena, CA 91125, USA}

\preprint{CALT-TH-2021-025}

\begin{abstract}
	We revisit the calculation of bosonic dark matter absorption via electronic excitations. 
	Working in an effective field theory framework and consistently taking into account in-medium effects, we clarify the relation between dark matter and photon absorption. 
    As is well-known, for vector (dark photon) and pseudoscalar (axion-like particle) dark matter, the absorption rates can be simply related to the target material's optical properties. 
	However, this is not the case for scalar dark matter, where the dominant contribution comes from a different operator than the one contributing to photon absorption, which is formally next-to-leading-order and does not suffer from in-medium screening. 
	It is therefore imperative to have reliable first-principles numerical calculations and/or semi-analytic modeling in order to predict the detection rate. 
	We present updated sensitivity projections for semiconductor crystal and superconductor targets for ongoing and proposed direct detection experiments.
\end{abstract}

\maketitle
\newpage
\tableofcontents
\newpage

\section{Introduction}
\label{sec:intro}

Uncovering the nature of cosmic dark matter (DM) remains one of the major goals in particle physics. 
Recent advances in low-threshold detectors (\eg\ skipper charge-coupled devices~\cite{Tiffenberg:2017aac}, transition edge sensors~\cite{Pyle:2015pya,Rothe:2018bnc,Fink:2020noh,tesseract}, microwave kinetic inductance detectors~\cite{Colantoni:2020cet} and quantum evaporation of helium atoms~\cite{Maris:2017xvi}) coupled with new theoretical investigations of various small-gap materials (\eg\ $\mathcal{O}(\text{eV})$-gap semiconductor crystals~\cite{Essig:2011nj,Graham:2012su,Essig:2012yx,Lee:2015qva,Essig:2015cda,Derenzo:2016fse,Hochberg:2016sqx,Bloch:2016sjj,Kurinsky:2019pgb,Trickle:2019nya, Griffin:2019mvc,Griffin:2020lgd,Du:2020ldo, Griffin:2021znd}, $\mathcal{O}(\text{meV})$-gap superconductors~\cite{Hochberg:2015pha,Hochberg:2015fth,Hochberg:2016ajh} and Dirac materials~\cite{Hochberg:2017wce,Coskuner:2019odd,Geilhufe:2019ndy,Inzani:2020szg}) have opened up new possibilities in the pursuit of this goal, well beyond the scope of conventional searches based on nuclear recoils. 
In a direct detection experiment, DM may leave its trace not only via scattering off the target ions or electrons, but also via absorption if it is bosonic and has a mass that matches the difference between energy levels in the target system~\cite{Pospelov:2008jk,Dzuba:2010cw,An:2014twa,Hochberg:2016ajh,Hochberg:2016sqx,Bloch:2016sjj,Knapen:2016cue,Hochberg:2017wce,Arvanitaki:2017nhi,Knapen:2017ekk,Griffin:2018bjn,Kurinsky:2019pgb,Coskuner:2019odd,Geilhufe:2019ndy,Lawson:2019brd,Gelmini:2020kcu,Gelmini:2020xir,Bloch:2020uzh,Tan:2021nif}. 
In this work, we focus on processes where the absorption of a bosonic DM drives electronic excitations, \ie\ transitions between electronic states.

It has been widely appreciated that, for several well-motivated bosonic DM models, the absorption process is closely related to that of photon absorption, and the rate can be expressed in terms of the target material's optical properties, \ie\ the (complex) conductivity or dielectric function. 
In fact, most studies on DM absorption so far have utilized this feature to make rate predictions by simply rescaling optical data.  
This approach is obviously attractive because it saves the labor of first-principles calculations, which can be technically challenging or resource-intensive, and because one can often make quick comparisons between target materials based on existing data.

Nevertheless, this data-driven approach has important limitations. 
First of all, conductivity/dielectric data are not always readily available, especially for newly proposed, more exotic materials, in which case one has to resort to first-principles calculations and/or semi-analytic modeling (this is the case, \eg\ for Dirac materials studied in several recent works~\cite{Hochberg:2017wce,Coskuner:2019odd,Geilhufe:2019ndy,Inzani:2020szg}). 
Meanwhile, and more importantly, the question of whether DM absorption for a particular model can be simply related to photon absorption is a nontrivial one, and explicit calculations are needed to establish the answer.

It is the purpose of this work to revisit the calculation of DM absorption via electronic excitations. 
We critically examine the question above by carefully working out the matching between relativistic Lagrangians for DM-electron interactions and non-relativistic (NR) effective field theories (EFTs) (Sec.~\ref{sec:nr_matching}), and computing in-medium self-energies to fully account for mixing and screening effects (Sec.~\ref{sec:in-medium}). 
This is a slightly different strategy than several previous calculations: by matching onto a NR EFT from the beginning instead of taking the NR limit of a relativistic calculation in the end, the power counting relevant for the absorption process becomes more transparent; also, the cryogenic nature of direct detection experiments allows us to perform the in-medium calculation in the zero-temperature limit and avoid the complications of thermal field theory.
We will carry out the calculation for three widely-studied bosonic DM candidates: 
\begin{itemize}
	\item 
	Vector (\eg\ dark photon) DM, which can be produced, for example, by inflationary fluctuations~\cite{Graham:2015rva}, by parent particle decays or coherent oscillations after reheating~\cite{Agrawal:2018vin,Dror:2018pdh,Co:2018lka,Bastero-Gil:2018uel}, or from a network of cosmic strings~\cite{Long:2019lwl}. 
	In this case, since the DM couples to electrons via the same vector current $\bar\psi\gamma^\mu\psi$ as the photon does, its absorption rate is trivially a rescaling of the photon absorption rate. 
	\item 
	Pseudoscalar (\eg\ axion-like particle) DM, which can be produced, for example, via the misalignment mechanism~\cite{Preskill:1982cy,Abbott:1982af,Dine:1982ah}, from the decays of topological defects~\cite{Davis:1986xc, Gorghetto:2018myk, Gorghetto:2020qws}, or by a variety of other mechanisms (see \eg\ Refs.~\cite{Lyth:1991ub,Visinelli:2009kt,Co:2018mho,Co:2019jts,Co:2020dya}). 
	While not immediately obvious (since the DM couples to a different current, $\bar\psi i\gamma^5\psi$, than the photon does), it has been well-known that also in this case, there is a simple relation between DM and photon absorption~\cite{Pospelov:2008jk}. 
	We will recover this result in the NR EFT calculation. 
	It is worth noting that the dominant contribution to NR pseudoscalar DM absorption actually comes from an operator generated at the next-to-leading order (NLO) in the $1/m_e$ expansion, because the leading order (LO) operator suffers a suppression by the DM's momentum $q$.
	\item 
	Scalar DM, which can be produced via mechanisms similar to pseudoscalar DM mentioned above.
	It couples to the scalar current $\bar\psi\psi$, which at LO coincides with the temporal component of the vector current $\bar\psi\gamma^0\psi$. 
	However, as we will see, the LO operator gives a $q$-suppressed contribution and, as in the pseudoscalar case, the rate is dominated by a NLO operator. 
	Importantly, this NLO operator has a different structure than the photon coupling, and its contribution cannot be simply related to photon absorption, invalidating the data-driven approach.
\end{itemize}
We make the statements above on the scalar and pseudoscalar DM more concrete in Table~\ref{tab:summary}. 

\begin{table}[t]
	\centering
	\begin{tabular}{c|cc|cc}
		\hline
		DM type & \multicolumn{2}{c|}{Scalar ($\phi\,\bar\psi\psi$)} & \multicolumn{2}{c}{Pseudoscalar ($\phi\,\bar\psi i\gamma^5\psi$)} \Tstrut\Bstrut\\
		\hline
		NR operators
		& \;\;$\phi\,\psi_+^\dagger\psi_+$ & $+$ \;\underline{$\frac{1}{8m_e^2}\,\phi\,\psi_+^\dagger\overleftrightarrow{\nabla}^2\psi_+$}\;\; & 
		\; $-\frac{1}{2m_e}\,(\nabla\phi)\cdot(\psi_+^\dagger \vect{\Sigma}\,\psi_+)$ & $+$\, \underline{$\frac{i}{4m_e^2} \,(\partial_t\phi)(\psi_+^\dagger \vect{\Sigma}\cdot\overleftrightarrow{\nabla}\psi_+)$} \;
		\Tstrut\Bstrut\\
		\;\;Related to dielectric?\;\;
		&
        \textcolor{ForestGreen}{$\checkmark$} & \textcolor{Red}{\xmark} & 
        \textcolor{ForestGreen}{$\checkmark$} & \textcolor{ForestGreen}{$\checkmark$}  \Tstrut\Bstrut\\
		\hline
	\end{tabular}
	\caption{
		\label{tab:summary}
		Summary of results for scalar and pseudoscalar DM $\phi$ coupling to electron $\psi$. 
		The effective operators at LO and NLO in the NR ($1/m_e$) expansion are shown in the second row. 
		In both cases, the NLO operator (underlined) gives the dominant contribution to DM absorption. 
		Importantly, the dominant contribution in the scalar case is not directly related to the target material's conductivity/dielectric function.
		See Secs.~\ref{sec:nr_matching} and \ref{sec:in-medium} for details.
	}
\end{table}

The fact that the DM absorption rate is not always relatable to the target material's optical properties highlights the necessity to go beyond the conventional data-driven approach. (The same can be said for DM scattering, for which the data-driven approach based on the dielectric function that has been advocated recently~\cite{Hochberg:2021pkt,Knapen:2021run,Knapen:2021bwg} covers only a limited set of DM interactions.) 
In this work, we consider two types of targets:
\begin{itemize}
	\item Semiconductor crystals with $\mathcal{O}(\text{eV})$ gaps (Sec.~\ref{sec:semiconductor}), focusing on silicon (Si) and germanium (Ge) that are in use in current experiments (DAMIC~\cite{deMelloNeto:2015mca,Aguilar-Arevalo:2019wdi,Settimo:2020cbq}, EDELWEISS~\cite{Armengaud:2018cuy, Armengaud:2019kfj, Arnaud:2020svb}, SENSEI~\cite{Crisler:2018gci,Abramoff:2019dfb,Barak:2020fql}, SuperCDMS~\cite{Agnese:2014aze, Agnese:2015nto, Agnese:2016cpb, Agnese:2017jvy, Agnese:2018col, Agnese:2018gze, Amaral:2020ryn}). 
        We compute DM absorption rates using first-principles density functional theory (DFT) calculations of electronic band structures and wave functions, which are now publicly available~\cite{sinead_m_griffin_2021_4735777}. 
	The numerical calculation builds upon the \texttt{EXCEED-DM} framework~\cite{Griffin:2021znd} and we publish the  ``absorption'' module of the program together with this work~\cite{exceed_dm_collaboration_2021_5009167}. 
	\item Conventional (BCS) superconductors with $\mathcal{O}(\text{meV})$ gaps (Sec.~\ref{sec:superconductor}), focusing on aluminum (Al) that has been proposed for direct detection~\cite{Hochberg:2015pha,Hochberg:2015fth,Hochberg:2016ajh}. We compute DM absorption rates by semi-analytically modeling the electronic states near the Fermi surface, largely following Refs.~\cite{Hochberg:2015pha,Hochberg:2015fth,Hochberg:2016ajh}.
\end{itemize}
For all the materials under study, we find good agreement between our theoretical calculation and the data-driven approach for the DM models where both are valid, \ie\ vector and pseudoscalar DM. 
This serves as an important validation of our calculations.
In the case of scalar DM, we show explicitly how the data-driven approach fails to reproduce the leading contribution, and present our calculated sensitivity projections.
In particular, for Al superconductor, our revised projected reach is much more optimistic than that found in Ref.~\cite{Gelmini:2020xir}, although somewhat weaker than the original estimate in Ref.~\cite{Hochberg:2016ajh}.

\section{Dark Matter Couplings to Non-relativistic Electrons}
\label{sec:nr_matching}

Since electrons in a detector are non-relativistic, it is convenient to perform the DM absorption calculation in the framework of NR EFT (see \eg\ Refs.~\cite{Rothstein:2003mp,Penco:2020kvy} for reviews). 
In this section, we work through the procedure of matching a relativistic theory of DM-electron interactions onto effective operators involving the NR electron field. 
The total Lagrangian of interest is
\begin{equation}
\mathcal{L} = \mathcal{L}_\psi + \mathcal{L}_\phi +\mathcal{L}_\text{int}\,.
\end{equation}
Here $\mathcal{L}_\psi$ is the Standard Model part that includes the electron $\psi$ coupling to electromagnetism,
\begin{equation}
\mathcal{L}_\psi = \bar\psi\,\bigl[i\gamma^\mu (\partial_\mu +i e A_\mu) -m_e\bigr]\psi \,,
\end{equation}
$\mathcal{L}_\phi$ contains the standard kinetic and mass terms of the DM field $\phi$, and we consider the following DM-electron interactions:
\begin{equation}
\mathcal{L}_\text{int} = 
\begin{cases}
\;g\phi \bar\psi \psi  & \text{(scalar DM,\; $g = d_{\phi ee} \frac{\sqrt{4\pi}\, m_e}{M_\text{Pl}}$)}\,, \\
\;g\phi \bar\psi i\gamma^5 \psi \simeq -\frac{g}{2m_e} (\partial_\mu\phi) (\bar\psi\gamma^\mu\gamma^5\psi) & \text{(pseudoscalar DM,\; $g=g_{aee}$)}\,, \\
\;g\phi_\mu \bar\psi \gamma^\mu \psi & \text{(vector DM,\; $g=\kappa e$)}\,,
\end{cases}
\label{eq:Lint}
\end{equation}
where we have also indicated the relation between the coupling $g$ and commonly adopted parameters $d_{\phi ee}$, $g_{aee}$, $\kappa$ in the literature. 
Note that there are two equivalent ways of writing the pseudoscalar coupling that are related by a field redefinition and integration by parts (IBP). 

Let us first consider $\mathcal{L}_\psi$. Writing the electron field in the relativistic theory as
\begin{equation}
\psi (\vect{x},\, t) = e^{-im_et}\, \psi_\text{NR}(\vect{x},\,t)\,.
\end{equation}
We obtain
\begin{equation}
\mathcal{L}_\psi 
= \psi^\dagger_\text{NR} \Bigl[ i\partial_t -eA_0 +i\gamma^0\vect{\gamma}\cdot(\nabla-ie\vect{A}) +\left(1-\gamma^0\right) m_e \Bigr] \psi_\text{NR} \,.
\label{eq:Lpsi}
\end{equation}
We now define projection operators
\begin{equation}
P_\pm \equiv \frac{1}{2}\left( 1\pm \gamma^0\right) ,
\end{equation}
which satisfy $P_\pm^2 = P_\pm$, $P_+P_-=P_-P_+ =0$ and $(P_\pm)^\dagger=P_\pm$. 
By using $P_\pm\gamma^0 = \gamma^0 P_\pm = \pm P_\pm$ and $P_\pm \gamma^i = \gamma^i P_\mp$, we can rewrite Eq.~\eqref{eq:Lpsi} as
\begin{equation}
\mathcal{L}_\psi = \psi_+^\dagger (i\partial_t -eA_0)\, \psi_+
+ \psi_-^\dagger (i\partial_t -eA_0 +2m_e)\, \psi_-
+\psi_+^\dagger \,i\vect{\gamma}\cdot(\nabla-ie\vect{A})\,\psi_-
-\psi_-^\dagger \,i\vect{\gamma}\cdot(\nabla-ie\vect{A})\,\psi_+ \,,
\end{equation}
where $\psi_\pm \equiv P_\pm \psi_\text{NR}$ (thus $\psi_\text{NR} = \psi_+ +\psi_-$). 
Integrating out the heavy field $\psi_-$ at tree level by solving its equation of motion (EOM),
\begin{equation}
\psi_- = \frac{1}{2m_e +i\partial_t -eA_0} \, i\vect{\gamma}\cdot (\nabla-ie\vect{A})\psi_+ \,,
\label{eq:eom}
\end{equation}
we arrive at the EFT for $\psi_+$:
\begin{align}
\mathcal{L}_\psi^\text{eff} =&\; \psi_+^\dagger \left[ i\partial_t -eA_0 
-\vect{\gamma}\cdot(\nabla-ie\vect{A})\, \frac{1}{2m_e+i\partial_t-eA_0}\, \vect{\gamma}\cdot(\nabla-ie\vect{A})\right] \psi_+ 
\nonumber\\
=&\; \psi_+^\dagger \left[ i\partial_t -eA_0 + \frac{(\nabla-ie\vect{A})^2}{2m_e} 
+(\nabla\times\vect{A})\cdot \frac{e\,\vect{\Sigma}}{2m_e}
-\frac{i}{4m_e^2}(\nabla-ie\vect{A})\cdot \partial_t\, (\nabla-ie\vect{A})
+\dots\right] \psi_+ 
\label{eq:L_psi_eff}
\end{align}
where we have used
\begin{equation}
\gamma^i\gamma^j = -\delta^{ij} -i\epsilon^{ijk}\Sigma^{k} \,,\qquad
\vect{\Sigma}\equiv \begin{pmatrix}
\vect{\sigma} & 0 \\ 0 & \vect{\sigma}
\end{pmatrix} .
\end{equation}
We can readily identify the first four terms in Eq.~\eqref{eq:L_psi_eff}, which come from LO in the $1/m_e$ expansion, as the familiar electromagnetic interactions as in NR quantum mechanics. 
There are several operators at NLO in the $1/m_e$ expansion, of which we have only written out the one involving $\partial_t$. 
This is the last term in Eq.~\eqref{eq:L_psi_eff}, and is the only NLO term that will be relevant in what follows. 
Importantly, it gives a tree-level contribution to the wave function renormalization of the $\psi_+$ field. 
In NR EFT calculations, it is often convenient to adopt an operator basis where temporal derivatives in the quadratic part of the Lagrangian have been traded for spatial derivatives, so as to eliminate any non-trivial wave function renormalization factors at tree level. 
The field redefinition needed to go into this basis, at the order we are working here, is
\begin{equation}
\psi_+ = \left[1 -\frac{1}{8m_e^2}\,\bigl(\vect{\gamma}\cdot(\nabla-ie\vect{A})\bigr)^2\right] \hat\psi_+ \,.
\label{eq:field_redef}
\end{equation}
This field redefinition does not change the LO Lagrangian (the first four terms in Eq.~\eqref{eq:L_psi_eff}), but replaces the last term in Eq.~\eqref{eq:L_psi_eff} by NLO operators that do not contain $\partial_t$ (and hence do not contribute to the wave function renormalization of $\hat\psi_+$). 
We will not need the NLO operators for electron couplings to vector fields (photon and dark photon),\footnote{As a side remark, in the special case of an electrostatic potential, $A_0=\Phi(\vect{x})$, $\vect{A}=\vect{0}$, one can check that keeping all the NLO terms reproduces the familiar fine structure correction in NR quantum mechanics: $\mathcal{L}_\psi^\text{eff,NLO} =\hat\psi_+^\dagger\, \frac{\nabla^4}{8m_e^3}\, \hat\psi_+
	-\frac{e}{8m_e^2}\, (\nabla^2\Phi)\, \hat\psi_+^\dagger\hat\psi_+ 
	-\frac{ie}{8m_e^2} \,(\nabla \Phi) \cdot \bigl(\hat\psi_+^\dagger\, \vect{\Sigma}\times \overleftrightarrow{\nabla} \hat\psi_+\bigr)$, where the three terms are the relativistic kinetic energy correction, the Darwin term and spin-orbit coupling, respectively.} but the field redefinition in Eq.~\eqref{eq:field_redef} that modifies the NLO Lagrangian will be important in the cases of scalar and pseudoscalar DM.

We are interested in the case where the photon field $A^\mu$ consists of an electrostatic background $\Phi$ and quantum fluctuations $\mathcal{A}^\mu$:
\begin{equation}
A_0 (\vect{x},t) = \Phi(\vect{x}) + \mathcal{A}_0 (\vect{x},t) \,,\qquad
\vect{A} (\vect{x},t) = \vect{\mathcal{A}}(\vect{x},t) \,.
\label{eq:A_bkg}
\end{equation}
The normalized NR field $\hat\psi_+$ can be expanded in energy eigenstates of the NR Schr\"odinger equation:
\begin{equation}
\hat\psi_+ (\vect{x}, t) = \sum_{I,s} \,\hat c_{I,s}\, e^{-i\varepsilon_It}\, \Psi_I(\vect{x}) \;\frac{1}{\sqrt{2}}
\begin{pmatrix}
\xi_s \\ \xi_s
\end{pmatrix}\,,
\end{equation}
where $\hat c_{I,s}$ are annihilation operators for NR electrons, and
\begin{equation}
\left( -\frac{\nabla^2}{2m_e} -e\,\Phi(\vect{x}) \right) \Psi_I(\vect{x}) = \varepsilon_I \Psi_I(\vect{x}) \,,\qquad
\xi_+ = \begin{pmatrix}
1 \\ 0
\end{pmatrix} \,,\qquad
\xi_- = \begin{pmatrix}
0 \\ 1
\end{pmatrix}\,.
\end{equation}
Note that the form of the background field in Eq.~\eqref{eq:A_bkg} assumes negligible spin-orbit coupling, in which case the two spin states $s=\pm$ for a given $I$ are degenerate.
From Eqs.~\eqref{eq:L_psi_eff} and \eqref{eq:A_bkg}, we can also deduce the electron's coupling to photon quanta $\mathcal{A}^\mu$ at LO in the NR EFT:
\begin{equation}
\mathcal{L}_{\psi\mathcal{A}}^\text{eff} = -e\,\mathcal{A}_0\, \hat\psi_+^\dagger \hat\psi_+ 
-\frac{ie}{2m_e} \,\vect{\mathcal{A}} \cdot \left(\hat\psi_+^\dagger \overleftrightarrow{\nabla} \hat\psi_+\right)
+\frac{e}{2m_e}\,(\nabla\times\vect{\mathcal{A}}) \cdot \left(\hat\psi_+^\dagger \,\vect{\Sigma}\, \hat\psi_+ \right)
-\frac{e^2}{2m_e}\,\vect{\mathcal{A}}^2\,\hat\psi_+^\dagger \hat\psi_+
\,,
\label{eq:L_psiA_eff}
\end{equation}
where $\hat\psi_+^\dagger \overleftrightarrow{\nabla} \hat\psi_+ \equiv \hat\psi_+^\dagger\, (\nabla \hat\psi_+) -(\nabla\hat\psi_+^\dagger)\, \hat\psi_+ $.

Let us now move on to the DM-electron interaction $\mathcal{L}_\text{int}$. 
For vector DM, we can simply replace $e\,\mathcal{A}^\mu\to e\,\mathcal{A}^\mu -g\,\phi^\mu$ in the derivation above, and obtain:
\begin{align}
\qquad
\mathcal{L}_\text{int}^\text{eff} =&\; g\,\phi_0\, \hat\psi_+^\dagger \hat\psi_+ 
+\frac{ig}{2m_e} \,\vect{\phi} \cdot \left(\hat\psi_+^\dagger \overleftrightarrow{\nabla} \hat\psi_+\right)
-\frac{g}{2m_e}\,(\nabla\times\vect{\phi}) \cdot \left(\hat\psi_+^\dagger \,\vect{\Sigma}\, \hat\psi_+ \right)&
\nonumber\\
&\;
+\frac{ge}{m_e}\,\vect{\phi}\cdot\vect{\mathcal{A}}\,\hat\psi_+^\dagger \hat\psi_+
-\frac{g^2}{2m_e}\,\vect{\phi}^2\,\hat\psi_+^\dagger \hat\psi_+
&\text{(vector DM).}
\end{align}
For the scalar and pseudoscalar cases, since $\mathcal{L}_\text{int}$ contains an operator that has a different structure than all the operators in $\mathcal{L}_\psi$, there is no such simple replacement. 
In principle, we should have included $\mathcal{L}_\text{int}$ when solving the EOM for $\psi_-$ in Eq.~\eqref{eq:eom}. 
However, if we are working at leading order in the DM-electron coupling $g$, it is sufficient to simply substitute Eq.~\eqref{eq:eom} into $\mathcal{L}_\text{int}$. 
We therefore obtain, at LO in the NR expansion:
\begin{equation}
\mathcal{L}_\text{int}^\text{eff,LO} = 
\begin{cases}
g\,\phi\, \hat\psi_+^\dagger \hat\psi_+ & \text{(scalar DM)}\,,\\
-\frac{g}{2m_e}\, (\nabla\phi)\cdot \hat\psi_+^\dagger \,\vect{\Sigma}\,\hat\psi_+  & \text{(pseudoscalar DM)}\,.
\end{cases}
\label{eq:L_int_LO}
\end{equation}
We now show that these LO terms are not sufficient to capture the dominant contributions to DM absorption. 
The point is that our NR EFT is an expansion in $\frac{\nabla}{m_e} \sim v_e$, and the power counting is such that momenta (and spatial derivatives) count as $m_e v_e$ and energies (and time derivatives) count as $m_e^{} v_e^2$.
For NR absorption, the energy deposition is $\omega \simeq m_\phi \sim m_e^{} v_e^2$. 
Meanwhile although the momentum transfer formally counts as $m_e v_e$, it is in fact much smaller: $q= m_\phi v_\phi \sim m_e^{} v_e^2 v_\phi \ll m_e v_e$, with $v_\phi\sim\mathcal{O}(10^{-3})$. 
Therefore, when the LO result contains factors of $q$, we need to work out the NLO terms and see if they may in fact dominate.

From Eq.~\eqref{eq:L_int_LO} it is clear that such $q$ suppression is indeed present in the pseudoscalar case.
It is perhaps less obvious that the scalar case also suffers a $q$ suppression, and its origin can be understood from charge conservation: the LO operator couples the scalar DM $\phi$ to the electron number density $\hat\psi_+^\dagger \hat\psi_+ = -\rho_e/e$ (with $\rho_e$ the charge density carried by the electron), whose matrix elements vanish in the $q\to0$ limit because $\rho_e = \vect{q}\cdot\vect{J}_e/\omega$; technically this is manifest via the orthogonality of initial and final state electron wave functions, as we will see later in the paper.\footnote{The same can be said for the $\phi_0$ component in the vector DM case. However, since $\phi_0$ couples exactly to the charge density even beyond LO, retaining higher order terms in the NR expansion does not remove the $q$ suppression.}
Therefore, in both scalar and pseudoscalar cases, we need to expand $\mathcal{L}_\text{int}$ up to NLO where there are several operators. 
Many of them will not be needed, though, because they are also $q$ suppressed or involve too many fields to contribute to the in-medium self-energies to be computed in the next section. 
Including only the unsuppressed operators at NLO that contain up to four fields, we have
\begin{equation}
\mathcal{L}_\text{int}^\text{eff} = 
\begin{cases}
g\,\phi\, \hat\psi_+^\dagger \hat\psi_+ 
+\frac{g}{8m_e^2}\,\phi \left(\hat\psi_+^\dagger \overleftrightarrow{\nabla}^2 \hat\psi_+ \right)
-\frac{ige}{2m_e^2}\,\phi\,\vect{\mathcal{A}}\cdot \left(\hat\psi_+^\dagger \overleftrightarrow{\nabla} \hat\psi_+\right)
& \text{(scalar DM)}\,,\\[8pt]
-\frac{g}{2m_e}\, (\nabla\phi)\cdot \hat\psi_+^\dagger \,\vect{\Sigma}\,\hat\psi_+  
+\frac{ig}{4m_e^2}\,(\partial_t\phi) \left(\hat\psi_+^\dagger\,\vect{\Sigma}\cdot \overleftrightarrow{\nabla} \,\hat\psi_+\right)
& \text{(pseudoscalar DM)}\,.
\end{cases}
\label{eq:L_int_eff}
\end{equation}
These results were already summarized in Table~\ref{tab:summary} (for brevity we dropped the hat on $\hat\psi_+$ and omitted the last operator in the scalar case in that table --- we will see that it gives vanishing contribution to DM absorption in an isotropic medium).
The second term in the scalar case, where the DM $\phi$ couples to $\hat\psi_+^\dagger \overleftrightarrow{\nabla}^2 \hat\psi_+ \equiv \hat\psi_+^\dagger (\nabla^2 \hat\psi_+) + (\nabla^2\hat\psi_+^\dagger) \,\hat\psi_+ -2\,(\nabla\hat\psi_+^\dagger)\cdot(\nabla\hat\psi_+)$, 
is obtained by combining the $\psi_-^\dagger\psi_-$ term from $\bar\psi\psi=\psi_+^\dagger\psi_+ -\psi_-^\dagger\psi_-$ (with $\psi_-$ replaced by its EOM solution Eq.~\eqref{eq:eom}) and additional terms from the field redefinition in Eq.~\eqref{eq:field_redef}. 
We will see in the next section that this operator gives the dominant contribution to scalar DM absorption. 
Pseudoscalar DM absorption is likewise dominated by the NLO operator $(\partial_t\phi) \bigl(\hat\psi_+^\dagger\,\vect{\Sigma}\cdot \overleftrightarrow{\nabla} \,\hat\psi_+\bigr)$.\footnote{Technically, the electron fields in the two equivalent expressions of the pseudoscalar coupling, $g\phi \bar\psi i\gamma^5 \psi$ and $-\frac{g}{2m_e} (\partial_\mu\phi) (\bar\psi\gamma^\mu\gamma^5\psi)$, are not the same, but are related by a field redefinition. If one derives the NR EFT starting from $-\frac{g}{2m_e} (\partial_\mu\phi) (\bar\psi\gamma^\mu\gamma^5\psi)$, this NLO operator is obtained directly from its $\mu=0$ component. On the other hand, if one derives the NR EFT from $g\phi \bar\psi i\gamma^5 \psi$, a further field redefinition is needed to eliminate operators involving the background electrostatic potential $\Phi$ and arrive at the same operator coefficient shown in Eq.~\eqref{eq:L_int_eff}.}

\section{In-medium Self-energies and Absorption Rates}
\label{sec:in-medium}

We now use the NR EFT derived in the previous section to compute DM absorption rates.
Generally, the absorption rate of a state can be derived from the imaginary part of its self-energy. 
In a medium, care must be taken because of mixing effects. 
If the DM $\phi$ mixes with a SM state $A$ in the medium (generalization to the case of mixing with multiple states is straightforward) then the self-energy matrix has to diagonalized to find the in-medium eigenstates:
\begin{equation}
\begin{pmatrix}
m_\phi^2 +\Pi_{\phi\phi} & \Pi_{\phi A} \\
\Pi_{A\phi} & \Pi_{AA}
\end{pmatrix}
\;\to\;
\begin{pmatrix}
m_\phi^2 + \Pi_{\hat\phi\hat\phi} & 0 \\
0 & \Pi_{\hat A \hat A}
\end{pmatrix} \,,
\end{equation}
where $\Pi_{\phi\phi}\sim\mathcal{O}(g^2)$, $\Pi_{\phi A}, \Pi_{A\phi} \sim\mathcal{O}(g)$. 
For a 4-momentum $Q^\mu = (\omega, \vect{q})$, we have $\Pi_{\phi A}(Q) =\Pi_{A\phi}(-Q)$. 
Simple algebra shows that to $\mathcal{O}(g^2)$,
\begin{equation}
\Pi_{\hat\phi\hat\phi} 
\simeq \Pi_{\phi\phi} +\frac{\Pi_{\phi A}\Pi_{A\phi}}{m_\phi^2 -\Pi_{AA}} \,.
\end{equation}
The DM absorption rate is then derived from the imaginary part of the eigenvalue corresponding to the DM-like state, $\hat\phi$:
\begin{equation}
\Gamma_\text{abs}^\phi = -\frac{Z_{\hat\phi}}{\omega} \,\text{Im}\,\Pi_{\hat\phi\hat\phi} 
\simeq -\frac{1}{\omega} \,\text{Im} \left( \Pi_{\phi\phi} +\frac{\Pi_{\phi A}\Pi_{A\phi}}{m_\phi^2 -\Pi_{AA}}\right),
\end{equation}
where the wave function renormalization $Z_{\hat\phi} = \Bigl(1-\frac{d\,\text{Re}\,\Pi_{\hat\phi\hat\phi}}{d\,\omega^2}\Bigr)^{-1} = 1+\mathcal{O}(g^2)$ has been approximated as unity. 
The total rate per unit target mass is given by
\begin{equation}
R = \frac{\rho_\phi}{\rho_T^{}}\,\frac{1}{\omega}\,\Gamma_\text{abs}^\phi
= -\frac{\rho_\phi}{\rho_T^{}}\,\frac{1}{\omega^2}\,\text{Im} \left( \Pi_{\phi\phi} +\frac{\Pi_{\phi A}\Pi_{A\phi}}{m_\phi^2 -\Pi_{AA}}\right)\,,
\label{eq:R}
\end{equation}
where $\rho_T^{}$ is the target's mass density, and $\rho_\phi=0.4\,$GeV/cm$^3$ is the local DM energy density. 
For non-relativistic DM, $\omega \simeq m_\phi$, and $\rho_\phi\simeq\frac{1}{2}m_\phi^2\phi_0^2$ with the DM field amplitude defined by $\phi(\vect{x},t) = \phi_0\cos(\vect{q}\cdot\vect{x}-\omega t)$.

The calculation of self-energies generally involves two graph topologies:
\begin{align}
\parbox[c][60pt][c]{120pt}{\centering
	\begin{fmffile}{diags/se1-general}
	\begin{fmfgraph*}(100,40)
	\fmfleft{in}
	\fmfright{out}
	\fmf{photon,tension=2,label=$\overset{Q}{\longrightarrow}$,l.side=left,l.d=4pt}{in,v1}
	\fmf{photon,tension=2}{v2,out}
	\fmfpoly{smooth,shade,tension=0.25}{b1,b2,b3,b4}
	\fmf{fermion,tension=1}{b2,v1,b1}
	\fmf{fermion,tension=1}{b4,v2,b3}
	\fmfv{decor.shape=circle,decor.filled=30,decor.size=3thick,label={\footnotesize $\mathcal{O}_1$},label.angle=-90,l.d=8pt}{v1}
	\fmfv{decor.shape=circle,decor.filled=30,decor.size=3thick,label={\footnotesize $\mathcal{O}_2$},label.angle=-90,l.d=8pt}{v2}
	\end{fmfgraph*}
	\end{fmffile}
}
\equiv&\; -i\,\Pi_{\mathcal{O}_1,\mathcal{O}_2}(Q) = -i\,\Pi_{\mathcal{O}_2,\mathcal{O}_1}(-Q) \,,\label{eq:graph_topology_1}\\
\parbox[t][80pt][t]{120pt}{\centering
	\begin{fmffile}{diags/se2-general}
	\begin{fmfgraph*}(80,60)
	\fmfleft{in,l}
	\fmfright{out,r}
	\fmftop{t}
	\fmfbottom{b}
	\fmf{photon,tension=2,label=$\overset{Q}{\longrightarrow}\;\;$,l.side=left,l.d=4pt}{in,v1}
	\fmf{photon,tension=2}{v1,out}
	\fmfpoly{smooth,shade,tension=0.15}{b1,b2,b3}
	\fmf{phantom,tension=5}{t,b1}
	\fmf{phantom,tension=5}{b,v1}
	\fmf{fermion,tension=1}{v1,b2,b3,v1}
	\fmfv{decor.shape=circle,decor.filled=30,decor.size=3thick,label={\footnotesize $\mathcal{O}$},label.angle=-90,l.d=8pt}{v1}
	\end{fmfgraph*}
	\end{fmffile}
}
\equiv&\; -i\,\Pi'_{\mathcal{O}}(Q) \,,
\label{eq:graph_topology_2}
\end{align}
where a blob represents the sum of one-particle-irreducible (1PI) graphs. 
While we have drawn curly external lines for concreteness, they can each represent a scalar, pseudoscalar or vector.
The operators $\mathcal{O}_1$, $\mathcal{O}_2$, $\mathcal{O}$ that the external fields couple to may carry Lorentz indices, in which case $\Pi$ and $\Pi'$ inherit these indices.
We discuss the calculation of these self-energy diagrams in App.~\ref{app:self-energy}.

In the cases of interest here, $A$ represents one of the polarizations of the SM photon, and $\Pi_{AA}$ is directly related to the target's complex conductivity/dielectric function, as discussed further below. 
Since $\Pi_{AA}$ enters the absorption rate formula (Eq.~\eqref{eq:R}) as long as there is a nonzero mixing $\Pi_{\phi A}$, let us examine this quantity in more detail before specializing to each DM model. 
The photon self-energy tensor $\Pi^{\mu\nu}$ is defined such that the effective action contains
\begin{align}
\mathcal{S}_\text{eff}\supset&\; \frac{1}{2} \int d^4Q\,\Pi^{\mu\nu}(Q) \mathcal{A}_\mu(Q) \mathcal{A}_\nu(-Q) \nonumber\\
=&\; \frac{1}{2} \int d^4Q\,\Bigl[ \Pi^{00}(Q) \mathcal{A}_0(Q) \mathcal{A}_0(-Q) -2\,\Pi^{0j}(Q) \mathcal{A}_0(Q) \mathcal{A}^j(-Q) +\Pi^{ij}(Q) \mathcal{A}^i(Q) \mathcal{A}^j(-Q)\Bigr]\,,
\label{eq:S_eff_A}
\end{align}
where $\mathcal{A}^j$ ($j=1,2,3$) represent the three components of $\vect{\mathcal{A}}$. 
As usual, we compute $\Pi^{\mu\nu}$ from the sum of 1PI graphs. 
From Eq.~\eqref{eq:S_eff_A} it is clear that the sign convention here is such that $i\,\Pi^{00}$, $-i\,\Pi^{0j}$ and $i\,\Pi^{ij}$ are given by the sum of two-point 1PI graphs between $\mathcal{A}_0\mathcal{A}_0$, $\mathcal{A}_0\mathcal{A}^j$ and $\mathcal{A}^i\mathcal{A}^j$, respectively. 
From the photon-electron couplings in Eq.~\eqref{eq:L_psiA_eff}, we obtain $\Pi^{\mu\nu}$ in terms of $\Pi_{\mathcal{O}_1,\mathcal{O}_2}$ and $\Pi'_{\mathcal{O}}$ defined in Eq.~\eqref{eq:graph_topology_1} and \eqref{eq:graph_topology_2}:
\begin{align}
&\Pi^{00} = -e^2 \,\Pi_{\mathbb{1},\mathbb{1}}\,,\qquad
\Pi^{0j} = -e^2\,\Pi_{\mathbb{1},v^j} \,,\nonumber\\
&\Pi^{ij} = -e^2\,\Pi_{v^i, v^j} -\frac{e^2}{4m_e^2}(q^2\delta^{ij} -q^iq^j)\,\Pi_{\mathbb{1},\mathbb{1}} +\frac{e^2}{m_e}\,\delta^{ij}\Pi'_{\mathbb{1}}\,,
\label{eq:Pi_ij}
\end{align}
where the velocity operator $v^j$ is defined by
\begin{equation}
v^j \equiv -\frac{i\overleftrightarrow{\nabla}_j}{2m_e}\,.
\label{eq:v_op_def}
\end{equation}
Here and in what follows, we suppress the arguments $Q^\mu=(\omega, \vect{q})$ of self-energy functions where there is no confusion. 
To arrive at the expression of $\Pi^{ij}$ in Eq.~\eqref{eq:Pi_ij}, we have simplified the spin trace assuming the electron loop does not involve non-trivial spin structures; for example, 
\begin{equation}
\Pi_{\Sigma^i,\Sigma^j} = \frac{\text{tr}(\sigma^i\sigma^j)}{\text{tr}\,\mathbb{1}} \,\Pi_{\mathbb{1},\mathbb{1}} = \delta^{ij}\,\Pi_{\mathbb{1},\mathbb{1}}\,.
\label{eq:spin_tr}
\end{equation}
This assumption is obviously valid for one-loop self-energy diagrams. 
In the superconductor calculation in Sec.~\ref{sec:superconductor}, we will need two-loop self-energies with an internal phonon line; in that case the electron-phonon coupling is spin-independent, so the same simplification applies.

The photon self-energy satisfies the Ward identity $Q_\mu \Pi^{\mu\nu} = Q_\nu \Pi^{\mu\nu} = 0$.
From Eq.~\eqref{eq:Pi_ij} we see that this implies the following relations between $\Pi_{\mathbb{1},\mathbb{1}}$, $\Pi_{\mathbb{1},v^j}$, $\Pi_{v^i,v^j}$ and $\Pi'_{\mathbb{1}}$:
\begin{equation}
\omega\,\Pi_{\mathbb{1},\mathbb{1}} = q^j\,\Pi_{\mathbb{1},v^j}\,,\qquad
\omega\,\Pi_{\mathbb{1},v^j} = q^i \,\Pi_{v^i,v^j} -\frac{q^j}{m_e}\,\Pi'_{\mathbb{1}} \,.
\label{eq:relations-1}
\end{equation}
These relations can be explicitly checked with the one-loop-level expressions in Eqs.~\eqref{eq:diagram1} and \eqref{eq:diagram2}.

We can write $\Pi^{\mu\nu}$ in terms of its polarization components as follows:
\begin{equation}
\Pi^{\mu\nu} = -\sum_{\lambda,\lambda'=\pm,L} \Pi_{\lambda\lambda'}\, e_\lambda^\mu e_{\lambda'}^{\nu*}\,,
\label{eq:Pi_photon_general}
\end{equation}
where 
\begin{equation}
e_\pm^\mu = \frac{1}{\sqrt{2}}\, (0\,,\; \vect{\hat x} \pm i\vect{\hat y}) \,,\quad
e_L^\mu = \frac{1}{\sqrt{Q^2}}\, (q\,,\; \omega\vect{\hat z}) 
\end{equation}
for $Q^\mu = (\omega, \vect{q}) = (\omega, q\vect{\hat z})$.
These are the three photon polarizations in Lorenz gauge $Q_\mu e_\lambda^\mu=0$, and coincide with the three physical polarizations of a massive vector with $m^2=Q^2$. 

We will mostly focus on isotropic target materials in this work, and leave a discussion of the anisotropic case to App.~\ref{app:anisotropic}. 
For an isotropic medium, the $3\times 3$ matrix $\Pi_{\lambda\lambda'}$ is diagonal:
\begin{equation}
\begin{pmatrix}
\Pi_{++} & \Pi_{+-} & \Pi_{+L} \\
\Pi_{-+} & \Pi_{--} & \Pi_{-L} \\
\Pi_{L+} & \Pi_{L-} & \Pi_{LL} 
\end{pmatrix} 
\;\overset{\text{isotropic}}{\longrightarrow} \;
\begin{pmatrix}
\Pi_T & 0 & 0 \\
0 & \Pi_T & 0 \\
0 & 0 & \Pi_L
\end{pmatrix}
\end{equation}
where $\Pi_T$ and $\Pi_L$ are the transverse and longitudinal photon self-energies, respectively. 
The photon self-energy tensor $\Pi^{\mu\nu}$ therefore has the following form:
\begin{equation}
\Pi^{\mu\nu} \;\overset{\text{isotropic}}{\longrightarrow}\; -\Pi_T\, \bigl( e_+^\mu e_+^{\nu*} + e_-^\mu e_-^{\nu*}\bigr) -\Pi_L e_L^\mu e_L^{\nu*}
\;=\; -\begin{pmatrix}
\frac{q^2}{Q^2}\,\Pi_L & 0 & 0 & \frac{\omega q}{Q^2}\,\Pi_L \\
0 & \Pi_T & 0 & 0 \\
0 & 0 & \Pi_T & 0 \\
\frac{\omega q}{Q^2}\,\Pi_L & 0 & 0 & \frac{\omega^2}{Q^2}\,\Pi_L
\end{pmatrix} \,.
\label{eq:Pi_iso}
\end{equation}
From the linear response relation $J^\mu = -\Pi^{\mu\nu} A_\nu$\footnote{Strictly speaking, linear response theory relates $J^\mu$ and $A_\nu$ via the retarded Green's function $R^{\mu\nu}$, which differs from the time-ordered self-energy $\Pi^{\mu\nu}$ by the sign of the imaginary part at negative frequencies. This difference is however irrelevant for our calculations.} and Ohm's law $\vect{J} = \sigma \vect{E}=\sigma(i\omega\vect{\mathcal{A}} -i\vect{q} \mathcal{A}_0)$ we can relate $\Pi_T$ and $\Pi_L$ to the complex conductivity $\sigma$, which in turn is related to the complex dielectric $\varepsilon$ via $\sigma=i\omega(1-\varepsilon)$~\cite{Hochberg:2015fth,Coskuner:2019odd,Trickle:2019nya}: 
\begin{equation}
\Pi_T = -i\omega\sigma =\omega^2 (1-\varepsilon)  \,,\qquad 
\Pi_L = -i\omega Z_L^{-1} \sigma = Q^2 (1-\varepsilon)\,,
\end{equation}
where $Z_L = \omega^2/Q^2$. 
The real part of the conductivity $\sigma_1\equiv\text{Re}\,\sigma$ (the imaginary part of the dielectric) gives the photon absorption rate in medium:
\begin{equation}
\sigma_1 = \omega \,\text{Im}\,\varepsilon = -\frac{1}{\omega}\,\text{Im}\,\Pi_T  = -\frac{Z_L}{\omega}\,\text{Im}\,\Pi_L \,.
\end{equation}
We finally note that all the quantities introduced above -- the complex conductivity $\sigma$ and dielectric $\varepsilon$, and photon self-energies $\Pi_T$, $\Pi_L$ can be simply computed from $\Pi_{\mathbb{1},\mathbb{1}}$:
\begin{equation}
\varepsilon-1 = \frac{i\sigma}{\omega} = -\frac{\Pi_L}{Q^2} = -\frac{\Pi_T}{\omega^2} = -\frac{e^2}{q^2}\,\Pi_{\mathbb{1},\mathbb{1}} \,.
\label{eq:dielectric}
\end{equation}
With the photon part of the self-energy calculation completed, we now move on to consider self-energies involving the DM and compute DM absorption rates.

\subsection{Vector Absorption}

Since a vector DM couples to electrons in the same way as the photon, albeit with a coupling rescaled by $-g/e=-\kappa$, we have
\begin{equation}
\Pi_{\phi\phi}^{\mu\nu} = -\kappa\,\Pi_{\phi A}^{\mu\nu} 
= -\kappa\,\Pi_{A\phi}^{\mu\nu} 
= \kappa^2\,\Pi^{\mu\nu} \,.
\end{equation}
Each of the three polarizations of $\phi$ mixes with the corresponding polarization of the photon. 
Therefore, for the transverse (longitudinal) polarization, we simply set $\Pi_{\phi\phi} = -\kappa\, \Pi_{\phi A} = -\kappa\, \Pi_{A\phi} = \kappa^2\, \Pi_{AA}$ in Eq.~\eqref{eq:R}, with $\Pi_{AA} = \Pi_T$ ($\Pi_L$). 
As a result,
\begin{equation}
R_{T,L} = - \kappa^2\,\frac{\rho_\phi}{\rho_T^{}} \;\text{Im} \left(\frac{\Pi_{T,L}}{m_\phi^2 -\Pi_{T,L}}\right)
= -\kappa^2\,\frac{\rho_\phi}{\rho_T^{}} \,m_\phi^2\;\text{Im} \left(\frac{1}{m_\phi^2 -\Pi_{T,L}}\right) .
\end{equation}
The total absorption rate for an unpolarized vector DM is obtained by averaging over the three polarizations, $R = (2R_T + R_L)/3$. 
For NR absorption, we have $\omega^2\simeq Q^2 = m_\phi^2$, and $\Pi_T\simeq \Pi_L = m_\phi^2\,\frac{e^2}{q^2}\,\Pi_{\mathbb{1},\mathbb{1}}$ (see Eq.~\eqref{eq:dielectric}), so
\begin{equation}
R_\text{vector} = -\kappa^2\,\frac{\rho_\phi}{\rho_T^{}}\;\text{Im} \left(\frac{1}{1 -\frac{e^2}{q^2}\,\Pi_{\mathbb{1},\mathbb{1}}}\right) .
\label{eq:R_v}
\end{equation}
The rate is semi-independent of the momentum transfer (and hence the DM velocity) since $\Pi_{\mathbb{1},\mathbb{1}}$ generically scales as $q^2$.

The result can also be written in terms of the material's complex conductivity/dielectric:
\begin{equation}
R_\text{vector} =  -\kappa^2\,\frac{\rho_\phi}{\rho_T^{}} \;\text{Im}\left(\frac{1}{\varepsilon}\right)
=\kappa^2\,\frac{\rho_\phi}{\rho_T^{}}\,\frac{1}{|\varepsilon|^2}\,\frac{\sigma_1}{m_\phi}\,,
\label{eq:R_v_sigma}
\end{equation}
with $\varepsilon$, $\sigma_1$ evaluated at $\omega=m_\phi$, $q=0$. 
One may think of 
\begin{equation}
\frac{1}{|\varepsilon|^2} 
= \frac{m_\phi^4}{(m_\phi^2 -\text{Re}\,\Pi_L)^2 + (\text{Im}\,\Pi_L)^2}
\label{eq:screening_factor}
\end{equation}
as an in-medium screening factor, which suppresses the absorption rate compared to the obvious rescaling of photon absorption by $\kappa^2$~\cite{An:2013yua,An:2014twa,Hochberg:2016ajh,Hochberg:2016sqx}.

\subsection{Pseudoscalar Absorption}

A pseudoscalar does not mix with the photon due to parity mismatch,\footnote{The mixed self-energy $\Pi_{\phi A}^0$ ($\Pi_{\phi A}^j$) between $\phi$ and $A_0$ ($A^j$) has to be parity odd (even). For an isotropic target one must have $\Pi_{\phi A}^j\propto q^j$ while $\Pi_{\phi A}^0$ is a scalar function of $q^2$, so neither has the right parity if nonzero.} and we simply have $R = -\frac{\rho_\phi}{\rho_T^{}}\,\frac{1}{\omega^2}\,\text{Im} \,\Pi_{\phi\phi}$.
The pseudoscalar self-energy $\Pi_{\phi\phi}$ is defined such that the effective action contains
\begin{equation}
\mathcal{S}_\text{eff}\supset -\frac{1}{2} \int d^4Q\,\bigl[ m_\phi^2 +\Pi_{\phi\phi}(Q) \bigr] \,\phi(Q) \,\phi(-Q)\,.
\end{equation}
Therefore, $-i\,\Pi_{\phi\phi}$ is given by the sum of two-point 1PI graphs. 
From the pseudoscalar coupling in Eq.~\eqref{eq:L_int_eff}, we find, again after simplifying the spin trace as in Eq.~\eqref{eq:spin_tr}:
\begin{equation}
\text{Im}\,\Pi_{\phi\phi} = \frac{g^2}{4m_e^2} \,\text{Im}\,\Bigl[ q^2\,\Pi_{\mathbb{1},\mathbb{1}} 
-\omega q^j\,\bigl(\Pi_{\mathbb{1},v^j} +\Pi_{v^j,\mathbb{1}}\bigr)
+\omega^2\,\Pi_{v^j, v^j}\Bigr]
\label{eq:Pi_phiphi_ps}
\end{equation}
Comparing with Eq.~\eqref{eq:Pi_ij}, we see that $\text{Im}\,\Pi_{\phi\phi}$ for a pseudoscalar is closely related to the photon polarization $\Pi^{\mu\nu}$:
\begin{equation}\label{eq:R_ps_1}
\text{Im}\,\Pi_{\phi\phi} = -\frac{g^2}{e^2} \,\frac{1}{4m_e^2} \,\text{Im}\biggl[ q^2\,\Pi^{00} -\omega\, q^j\bigl(\Pi^{0j} +\Pi^{j0}\bigr) + \omega^2 \Pi^{jj} - q^2\, \frac{\omega^2}{2m_e^2}\,\Pi^{00} \biggr]\,.
\end{equation}
Note that the $\Pi'_{\mathbb{1}}$ term in $\Pi^{jj}$ is purely real and thus does not appear in the equation above. 
Also, since $\omega\ll m_e$, we can drop the last term. 
Writing $\Pi^{\mu\nu}$ in terms of $\Pi_T$ and $\Pi_L$ as in Eq.~\eqref{eq:Pi_iso} and setting $g=g_{aee}$, we find
\begin{equation}
R_\text{pseudoscalar} = -g_{aee}^2 \,\frac{\rho_\phi}{\rho_T^{}} \,\frac{1}{4m_e^2\omega^2}\, \frac{1}{e^2} \,\bigl(2\omega^2\,\text{Im}\,\Pi_T + m_\phi^2\,\text{Im}\,\Pi_L\bigr)\,.
\label{eq:R_ps_0}
\end{equation}
For NR absorption, $\omega^2\simeq Q^2 = m_\phi^2$, and $\Pi_T\simeq \Pi_L = e^2\,\frac{m_\phi^2}{q^2}\,\Pi_{\mathbb{1},\mathbb{1}}$ (see Eq.~\eqref{eq:dielectric}), and therefore, 
\begin{equation}
R_\text{pseudoscalar} 
= -g_{aee}^2\,\frac{\rho_\phi}{\rho_T^{}}\, \frac{3m_\phi^2}{4m_e^2}\, \frac{1}{q^2} \,\text{Im}\,\Pi_{\mathbb{1},\mathbb{1}}
\,.
\label{eq:R_ps}
\end{equation}
As in the vector DM case, the absorption rate can be written solely in terms of $\Pi_{\mathbb{1},\mathbb{1}}$; the other self-energies that appear in Eq.~\eqref{eq:Pi_phiphi_ps} have been traded for $\Pi_{\mathbb{1},\mathbb{1}}$ via the Ward identity. 
Also, analogous to the vector DM case, the rate is semi-independent of the DM velocity as $\Pi_{\mathbb{1},\mathbb{1}} \sim q^2$.
Note that the dominant contribution to pseudoscalar DM absorption comes from the last term in Eq.~\eqref{eq:Pi_phiphi_ps} that is proportional to $\omega^2\,\Pi_{v^i,v^j}$, which originates from the second (formally NLO) operator in Eq.~\eqref{eq:L_int_eff} (as underlined in Table~\ref{tab:summary}). 

We can further recast the pseudoscalar DM absorption rate in terms of the photon absorption rate $\sigma_1=\text{Re}\,\sigma = \omega\,\text{Im}\,\varepsilon$ and reproduce the standard result~\cite{Pospelov:2008jk,Hochberg:2016ajh,Hochberg:2016sqx, Bloch:2016sjj}:
\begin{equation}
R_\text{pseudoscalar} = \frac{g_{aee}^2}{e^2}\, \frac{\rho_\phi}{\rho_T^{}} \,\frac{3m_\phi \sigma_1}{4m_e^2}\,.
\label{eq:R_ps_sigma}
\end{equation}
We remark in passing that pseudoscalar absorption has also been studied in the context of solar axion detection; in that case, the relativistic kinematics $\omega\gg m_\phi$ means that the $\text{Im}\,\Pi_L$ term in Eq.~\eqref{eq:R_ps_0} is negligible, so the proportionality factor in Eq.~\eqref{eq:R_ps_sigma} is $\frac{1}{2}$ instead of $\frac{3}{4}$~\cite{Pospelov:2008jk,Derevianko:2010kz,Hochberg:2016sqx}.

\subsection{Scalar Absorption}

For scalar DM, we need to compute explicitly both $\text{Im}\,\Pi_{\phi\phi}$ and its mixing with the photon $\Pi_{\phi A}^\mu(Q)=\Pi_{A\phi}^\mu(-Q)$. 
These self-energies are defined such that
\begin{align}
\mathcal{S}_\text{eff}\supset&\; \int d^4Q\,\biggl[ -\frac{1}{2} \bigl( m_\phi^2 +\Pi_{\phi\phi}(Q) \bigr) \,\phi(Q)\,\phi(-Q) -\Pi_{\phi A}^\mu(Q)\,\phi(Q) \,\mathcal{A}_\mu(-Q)\biggr] \nonumber\\
=&\; \int d^4Q\,\biggl[ -\frac{1}{2} \bigl( m_\phi^2 +\Pi_{\phi\phi}(Q) \bigr) \,\phi(Q)\,\phi(-Q) -\Pi_{\phi A}^0(Q)\,\phi(Q) \,\mathcal{A}_0(-Q) +\Pi_{\phi A}^j(Q)\,\phi(Q) \,\mathcal{A}^j(-Q)\biggr]\,.
\end{align}
Therefore, $-i\,\Pi_{\phi\phi}$, $-i\,\Pi_{\phi A}^0$ and $i\,\Pi_{\phi A}^j$ are given by the sum of two-point 1PI graphs between $\phi\phi$, $\phi \mathcal{A}_0$ and $\phi\mathcal{A}^j$, respectively. 
From the scalar coupling in Eq.~\eqref{eq:L_int_eff} and photon coupling in Eq.~\eqref{eq:L_psiA_eff}, we find:
\begin{align}
\text{Im}\,\Pi_{\phi\phi} =&\; g^2\,\text{Im}\,\bigl(\Pi_{\mathbb{1},\mathbb{1}} -\Pi_{\mathbb{1},\bar v^2} -\Pi_{\bar v^2,\mathbb{1}} +\Pi_{\bar v^2,\bar v^2}\bigr) \,,\label{eq:Im_Pi_s}\\
\Pi_{\phi A}^0 =&\; -ge\,\bigl( \Pi_{\mathbb{1},\mathbb{1}} -\Pi_{\bar v^2,\mathbb{1}} \bigr) \,,\\
\Pi_{\phi A}^j =&\; -ge\,\biggl( \Pi_{\mathbb{1},v^j} -\Pi_{\bar v^2,v^j} +\frac{1}{m_e}\,\Pi'_{v^j}\biggr) \,,
\end{align}
where
\begin{equation}
\bar v^2 \equiv \frac{1}{2}\,v^j v^j = -\frac{\overleftrightarrow{\nabla}^2}{8m_e^2}\,.
\end{equation}
As in the photon case, the self-energies are related by the Ward identity $Q_\mu\Pi_{\phi A}^\mu=0$:
\begin{equation}
\omega\,\Pi_{\bar v^2,\mathbb{1}} = q^j\,\Pi_{\bar v^2,v^j} -\frac{q^j}{m_e}\,\Pi'_{v^j} \,,
\label{eq:relations-2}
\end{equation}
where we have used the first relation in Eq.~\eqref{eq:relations-1}. 
One can explicitly check that Eq.~\eqref{eq:relations-2} holds between the one-loop-level expressions for the self-energies in Eqs.~\eqref{eq:diagram1} and \eqref{eq:diagram2}.

For an isotropic medium, we must have $\Pi_{\phi A}^j\propto q^j$ because there is no special direction other than $\vect{q}$.\footnote{We note in passing that the $\Pi'_{v^j}$ term in $\Pi_{\phi A}^j$ is $\vect{q}$ independent and must therefore vanish in an isotropic medium. This is why we have omitted the $\phi\,\vect{\mathcal{A}}\cdot \bigl(\hat\psi_+^\dagger \overleftrightarrow{\nabla} \hat\psi_+\bigr)$ operator in Eq.~\eqref{eq:L_int_eff}, which only contributes to this term, from Table~\ref{tab:summary}.} 
So the mixing only involves the photon's longitudinal component. 
Therefore, $\Pi_{AA}$ in the rate formula Eq.~\eqref{eq:R} should be set to $\Pi_L=m_\phi^2\,\frac{e^2}{q^2}\,\Pi_{\mathbb{1},\mathbb{1}}$ (see Eq.~\eqref{eq:dielectric}), and $\Pi_{\phi A}$ should be set to
\begin{equation}
\Pi_{\phi L} = \Pi_{\phi A}^\mu e_{L\mu} = \frac{1}{q\sqrt{Q^2}} \bigl( q^2\,\Pi_{\phi A}^0 -\omega q^j\,\Pi_{\phi A}^j\bigr)
= -\frac{\sqrt{Q^2}}{q}\,\Pi_{\phi A}^0
= ge\,\frac{\sqrt{Q^2}}{q}\,\bigl(\Pi_{\mathbb{1},\mathbb{1}} -\Pi_{\bar v^2,\mathbb{1}}\bigr)\,,
\label{eq:Pi_phiL}
\end{equation}
where we have used the Ward identity to trade $q^j\,\Pi_{\phi A}^j$ for $\omega\,\Pi_{\phi A}^0$. 
Substituting the expressions for $\text{Im}\,\Pi_{\phi\phi}$, $\Pi_{\phi L}$ and $\Pi_L$ above into Eq.~\eqref{eq:R}, and applying the NR absorption kinematics $\omega^2\simeq Q^2=m_\phi^2$, we find
\begin{align}
R_\text{scalar} 
=&\; -d_{\phi ee}^2 \, \frac{4\pi m_e^2}{M_\text{Pl}^2}\,\frac{\rho_\phi}{\rho_T^{}}\,\frac{1}{m_\phi^2}\,\text{Im} \left[ \Pi_{\bar v^2, \bar v^2} +\frac{q^2}{e^2} \,\frac{\bigl(1-\frac{e^2}{q^2}\,\Pi_{\bar v^2,\mathbb{1}}\bigr)\bigl(1-\frac{e^2}{q^2}\,\Pi_{\mathbb{1},\bar v^2}\bigr)}{1-\frac{e^2}{q^2}\,\Pi_{\mathbb{1},\mathbb{1}}}\right],
\label{eq:R_s}
\end{align}
where we have used $\Pi_{L\phi}(Q)=\Pi_{\phi L}(-Q)$, $\Pi_{\bar v^2,\mathbb{1}}(-Q)=\Pi_{\mathbb{1},\bar v^2}(Q)$, and $g=d_{\phi ee}\,\frac{\sqrt{4\pi}\,m_e}{M_\text{Pl}}$.

We see that the result for scalar absorption, Eq.~\eqref{eq:R_s}, depends on $\Pi_{\bar v^2, \bar v^2}$, $\Pi_{\bar v^2,\mathbb{1}}$, $\Pi_{\mathbb{1},\bar v^2}$ in addition to $\Pi_{\mathbb{1},\mathbb{1}}$. 
If we had kept only the LO operator $\phi\, \hat\psi_+^\dagger\hat\psi_+$ in the calculation above, we would obtain Eq.~\eqref{eq:R_s} with $\Pi_{\bar v^2, \bar v^2}$, $\Pi_{\bar v^2,\mathbb{1}}$, $\Pi_{\mathbb{1},\bar v^2}$ set to zero, which coincides with $\frac{q^2}{m_\phi^2}$ times the vector DM absorption rate in Eq.~\eqref{eq:R_v}. 
Just as in the vector DM case, the contribution of the LO operator $\phi\, \hat\psi_+^\dagger\hat\psi_+$ to scalar DM absorption is screened due to in-medium mixing~\cite{Gelmini:2020xir}. 
However, the formally NLO operator $\phi\,\bigl(\hat\psi_+^\dagger \overleftrightarrow{\nabla}^2 \hat\psi_+ \bigr)$ introduces additional contributions via $\Pi_{\bar v^2, \bar v^2}$, $\Pi_{\bar v^2,\mathbb{1}}$, $\Pi_{\mathbb{1},\bar v^2}$. 
As we will see in the next two sections, generically $\Pi_{\mathbb{1}, \mathbb{1}}\,,\, \Pi_{\bar v^2,\mathbb{1}} \sim q^2$ while $\Pi_{\bar v^2, \bar v^2}\sim q^0$. 
It is thus clear from Eq.~\eqref{eq:R_s} that the absorption rate of a NR scalar DM is in fact dominated by the $\Pi_{\bar v^2, \bar v^2}$ term: 
\begin{equation}
R_\text{scalar} \simeq -d_{\phi ee}^2 \, \frac{4\pi m_e^2}{M_\text{Pl}^2}\,\frac{\rho_\phi}{\rho_T^{}}\,\frac{1}{m_\phi^2}\,\text{Im}\, \Pi_{\bar v^2, \bar v^2} \,.
\label{eq:R_s_approx}
\end{equation}
Importantly, this term (overlooked in several previous calculations of scalar DM absorption~\cite{Gelmini:2020xir,Bloch:2020uzh,Tan:2021nif}) is {\it not} directly proportional to the photon absorption rate and is unscreened. 
We emphasize that the suppression of LO operator's contribution is specific to the case of non-relativistic DM absorption, where $q\ll\omega$; for absorption of a relativistic scalar ($q\simeq\omega$) or scalar-mediated scattering ($q\gg\omega$), the LO operator $\phi\, \hat\psi_+^\dagger\hat\psi_+$ indeed gives the dominant contribution.
\\

To summarize, in this section we have derived DM absorption rates in terms of in-medium self-energies of the form $\Pi_{\mathcal{O}_1,\mathcal{O}_2}$, as defined in Eq.~\eqref{eq:graph_topology_1}. (Contributions from the other graph topology, Eq.~\eqref{eq:graph_topology_2}, have been eliminated using the Ward identity.) 
Both vector and pseudoscalar absorption involve a single self-energy function $\Pi_{\mathbb{1},\mathbb{1}}\propto\Pi_L$ (see Eqs.~\eqref{eq:R_v} and \eqref{eq:R_ps}), and the rates can be simply related to the (complex) conductivity/dielectric (see Eqs.~\eqref{eq:R_v_sigma} and \eqref{eq:R_ps_sigma}). 
In these cases, the data-driven approach based on the measured conductivity/dielectric is viable, and we can also use optical data to calibrate our theoretical calculations based on DFT or analytic modeling. 
On the other hand, for scalar DM absorption, additional self-energy functions $\Pi_{\bar v^2, \bar v^2}$, $\Pi_{\bar v^2,\mathbb{1}}$, $\Pi_{\mathbb{1},\bar v^2}$ enter (see Eq.~\eqref{eq:R_s}), and the rate is not directly related to photon absorption.
In this case, the data-driven approach fails and theoretical calculations are needed. 

\begin{figure}[t]
	\centering
    \includegraphics[width=\textwidth]{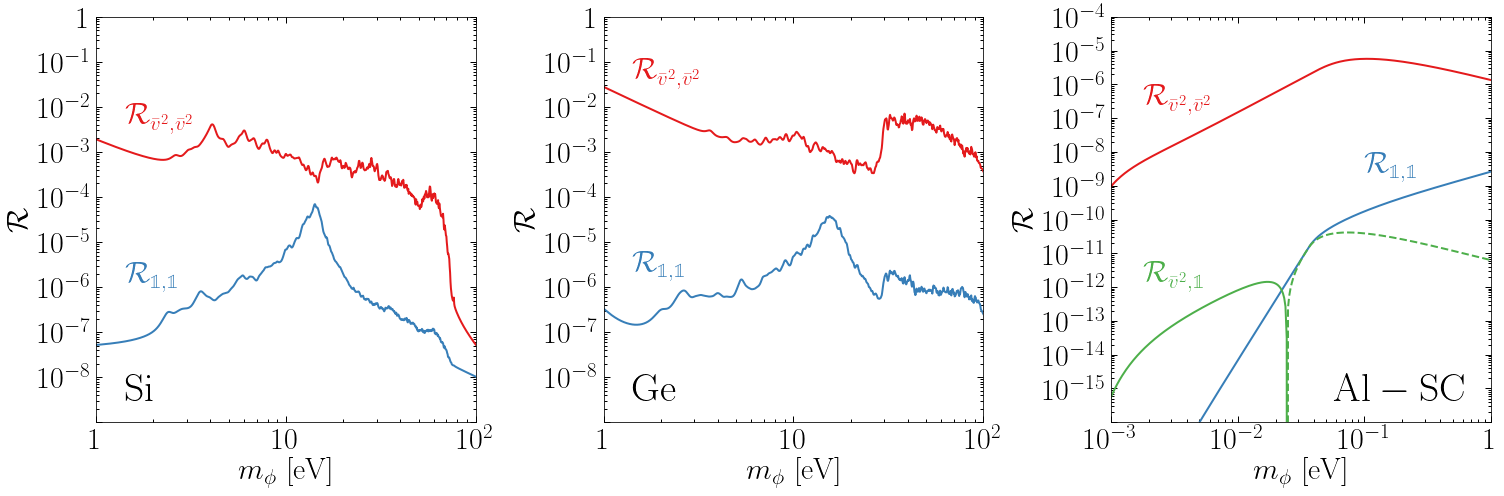}
	\caption{\label{fig:scalar_rate}
        Comparison between different terms contributing to the scalar DM absorption rate, defined in Eq.~\eqref{eq:R_split}, for Si, Ge and Al-SC targets assuming $q = 10^{-3} m_\phi$. 
        Dashed curves indicate negative values. 
        In all three targets we see that $\mathcal{R}_{\bar v^2, \bar v^2}$ dominates over the entire DM mass range considered. 
        This term comes from an NLO operator in the NR EFT (underlined in Table~\ref{tab:summary}) and cannot be directly related to the target's optical properties (\ie\ the complex conductivity/dielectric function). 
        For Si and Ge, the calculation of $\mathcal{R}_{\bar v^2, \mathbb{1}}$ is technically challenging as explained in Sec.~\ref{sec:semiconductor}; however, it is parameterically the same order in $q$ as $\mathcal{R}_{\mathbb{1}, \mathbb{1}}$ and therefore expected to be also subdominant compared to  $\mathcal{R}_{\bar v^2, \bar v^2}$.
	}
\end{figure}

\begin{figure}[t]
	\centering
	\includegraphics[width=0.49\textwidth]{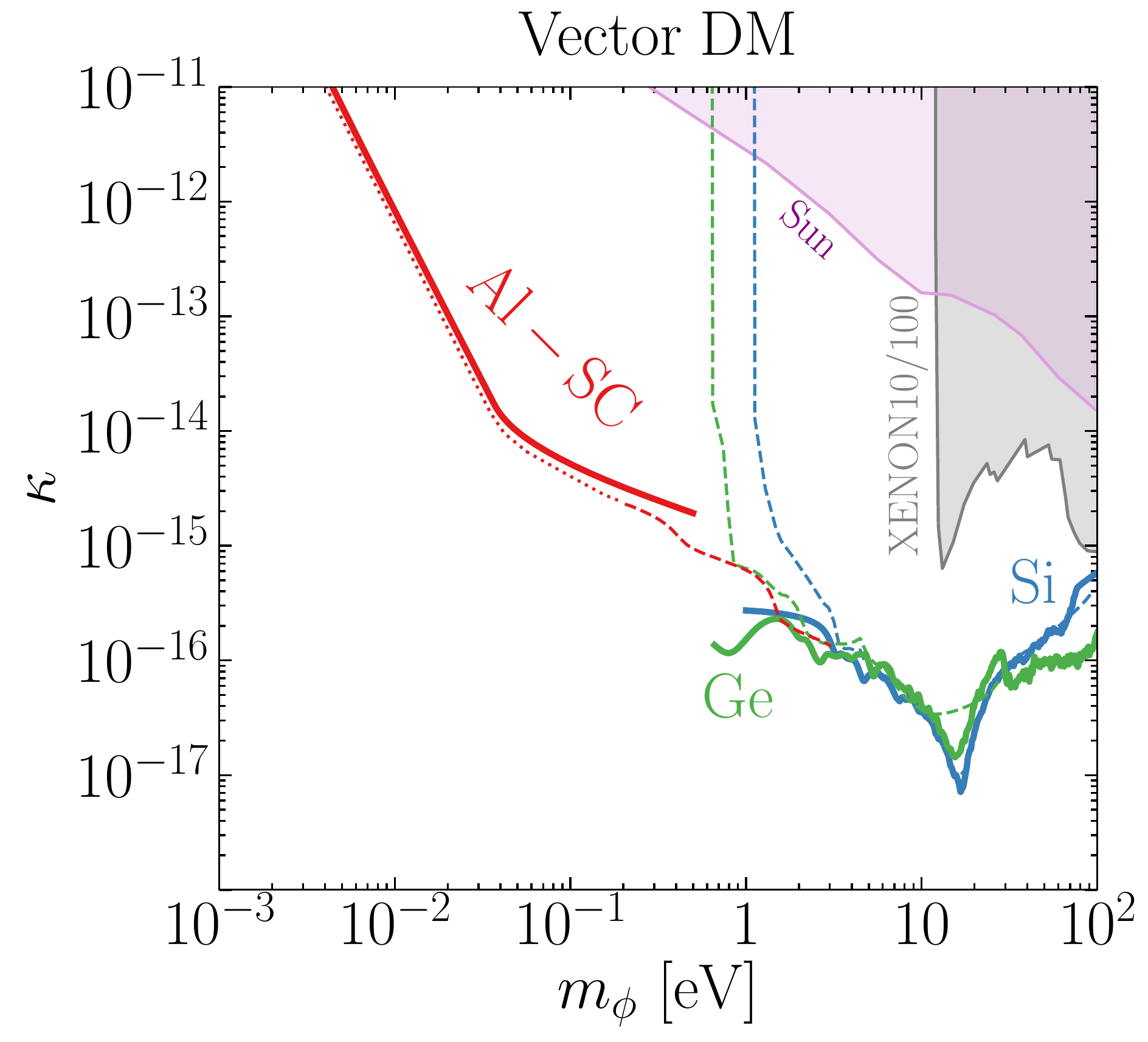}
	\includegraphics[width=0.49\textwidth]{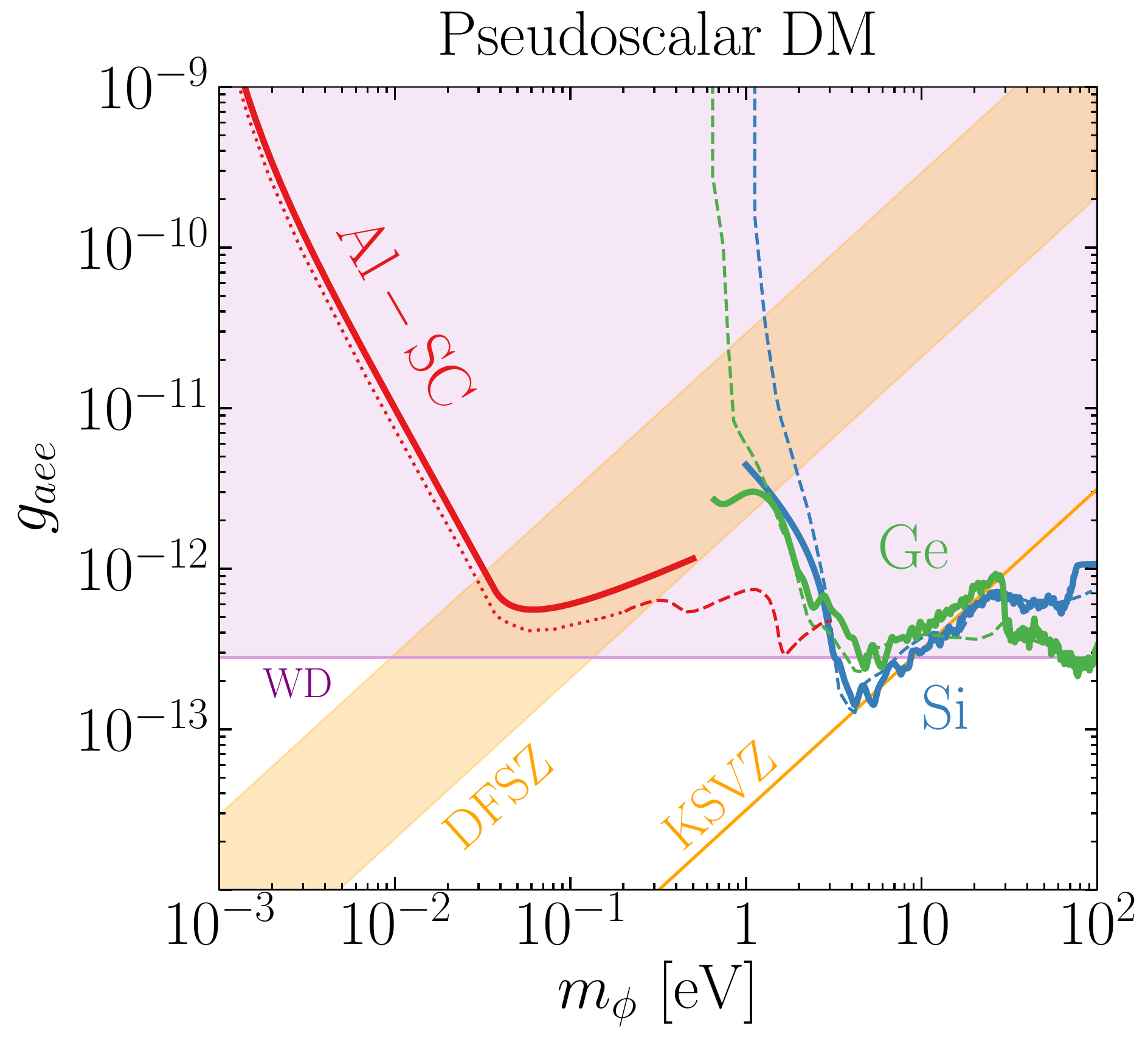}
	\caption{\label{fig:opt_compare_reach}
		Projected 95\% C.L. reach (3 events with no background) with semiconductor crystal (Si, Ge) and superconductor (Al-SC) targets for the vector and pseudoscalar DM models defined in Eq.~\eqref{eq:Lint}, assuming 1~kg-yr exposure. 
		We compare our theoretically calculated reach (solid) against the data-driven approach utilizing the target material's measured conductivity/dielectric~\cite{1985a,PhysRevB.12.5615} (dashed). 
		For Si and Ge, the data-driven approach was taken in previous works~\cite{Hochberg:2016sqx,Bloch:2016sjj}, with which we find good agreement. 
		For Al-SC, our theoretical calculation reproduces the results in Ref.~\cite{Hochberg:2016ajh} (dotted) up to the choice of overall normalization factor. 
		Also shown are existing direct detection limits from XENON10/100~\cite{Bloch:2016sjj}, stellar cooling constraints from the Sun (assuming St\"uckelberg mass for vector DM)~\cite{An:2013yfc} and white dwarfs (WD)~\cite{Bertolami:2014wua}, and pseudoscalar couplings corresponding to the QCD axion in KSVZ and DFSZ (for $0.28 \le \tan \beta \le 140$) models~\cite{Tanabashi:2018oca}.
	}
\end{figure}

\begin{figure}
    \centering
    \includegraphics[width=0.6\textwidth]{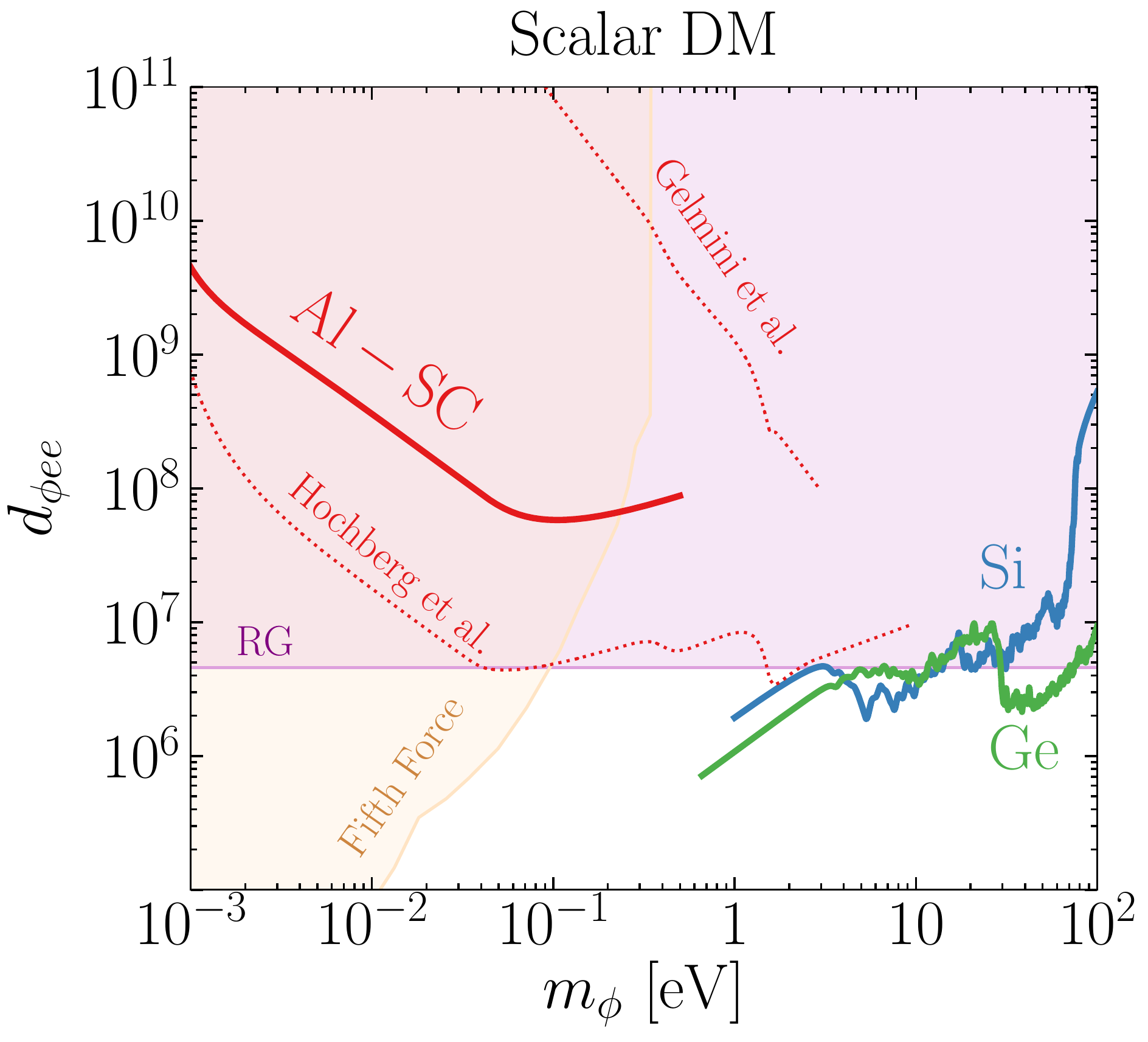}
    \caption{
    	Projected 95\% C.L. reach (3 events with no background) with semiconductor crystal (Si, Ge) and superconductor (Al-SC) targets, for the scalar DM model defined in Eq.~\eqref{eq:Lint}, assuming 1~kg-yr exposure. 
    	In contrast to the vector and pseudoscalar cases shown in Fig.~\ref{fig:opt_compare_reach}, the projections here cannot be derived from the target's optical properties. 
    	Differences compared to Hochberg {\it et al.}~\cite{Hochberg:2016ajh} and Gelmini {\it et al.}~\cite{Gelmini:2020xir} in the Al-SC case are discussed in detail in Sec.~\ref{sec:superconductor}. 
    	Also shown are existing constraints from fifth force~\cite{Adelberger:2003zx} and red giant (RG) cooling~\cite{Hardy:2016kme}. 
    }
    \label{fig:scalar_reach}
\end{figure}

In the next two sections, we compute the self-energies $\Pi_{\mathbb{1},\mathbb{1}}$, $\Pi_{\bar v^2, \bar v^2}$, $\Pi_{\bar v^2,\mathbb{1}}$, $\Pi_{\mathbb{1},\bar v^2}$ in crystal and superconductor targets, respectively, which then allow us to derive the absorption rates of vector, pseudoscalar and scalar DM in these targets. 
Our main results for Si, Ge and Al-superconductor (Al-SC) targets are collected in Figs.~\ref{fig:scalar_rate}, \ref{fig:opt_compare_reach} and \ref{fig:scalar_reach}. 
First, Fig.~\ref{fig:scalar_rate} confirms the dominance of the $\Pi_{\bar v^2, \bar v^2}$ term in the scalar DM absorption rate (\ie\ that Eq.~\eqref{eq:R_s} indeed simplifies to Eq.~\eqref{eq:R_s_approx}) by rewriting Eq.~\eqref{eq:R_s} as
\begin{equation}
R_\text{scalar}  = d_{\phi ee}^2 \, \frac{4\pi m_e^2}{M_\text{Pl}^2}\,\frac{\rho_\phi}{\rho_T^{}} \left( \mathcal{R}_{\bar v^2, \bar v^2} + \mathcal{R}_{\mathbb{1}, \mathbb{1}} + \mathcal{R}_{\bar v^2, \mathbb{1}} \right) ,
\label{eq:R_split}
\end{equation}
and comparing the sizes of the terms. 
Here $\mathcal{R}_{\bar v^2, \bar v^2} \equiv -\frac{1}{m_\phi^2}\,\text{Im}\,\Pi_{\bar v^2, \bar v^2}$, $\mathcal{R}_{\mathbb{1}, \mathbb{1}} \equiv -\frac{1}{m_\phi^2}\,\frac{q^2}{e^2}\,\text{Im}\,\biggl(\frac{1}{1-\frac{e^2}{q^2}\,\Pi_{\mathbb{1},\mathbb{1}}}\biggr)$, while the remaining terms define $\mathcal{R}_{\bar v^2, \mathbb{1}}$. 
Next, Fig.~\ref{fig:opt_compare_reach} shows the projected reach for the pseudoscalar and vector DM models, where we see good agreement between our theoretical calculations (solid curves) and rescaled optical data (dashed curves). Lastly, Fig.~\ref{fig:scalar_reach} shows our calculated reach for scalar DM and compares the Al-SC results with previous work~\cite{Hochberg:2016sqx, Gelmini:2020xir}.
These results will be discussed in detail in the following sections.

%%%
%%%

%%%

%%%

\section{Dark Matter Absorption in Crystals}
\label{sec:semiconductor}

In this section, we specialize to the case of crystal targets that are described by band theory. 
It suffices to compute the self-energies $\Pi_{\mathcal{O}_1,\mathcal{O}_2}$ at one-loop level, with $\mathcal{O}_{1,2}=\mathbb{1}, \bar v^2$. 
The result for general $\mathcal{O}_1$, $\mathcal{O}_2$ is given in Eq.~\eqref{eq:diagram1} in Appendix~\ref{app:self-energy}, and involves a sum over electronic states $I, I'$ that run in the loop. 
Since we assume the target is at zero temperature the occupation numbers $f_I$, $f_{I'}$ take values of either 1 or 0. 
Only pairs of states for which $f_{I'}-f_I\ne0$, \ie\ one is occupied and the other is unoccupied, contribute to the sum --- it is between these pairs of states that electronic transitions can happen. 

In the present case, the states are labeled by a band index $i$ and momentum $\vect{k}$ within the first Brillouin zone (1BZ), so we write $I=i, \vect{k}$, and $I'=i', \vect{k}'$. 
The wave functions have the Bloch form, which in real and momentum space read, respectively:
\begin{equation}
\Psi_{i,\vect{k}}(\vect{x}) = \frac{1}{\sqrt{V}}\,\sum_{\vect{G}} u_{i,\vect{k},\vect{G}}\,e^{i(\vect{k}+\vect{G})\cdot\vect{x}}\,,\qquad
\widetilde\Psi_{i,\vect{k}} (\vect{p}) = \sqrt{V} \,\sum_{\vect{G}} u_{i,\vect{k},\vect{G}} \,\delta_{\vect{p},\vect{k}+\vect{G}}\,,
\end{equation}
where the sum runs over all reciprocal lattice vectors $\vect{G}$. 
These are related by 
\begin{equation}
\Psi_{i,\vect{k}}(\vect{x}) = \int\frac{d^3p}{(2\pi)^3}\,\widetilde\Psi_{i,\vect{k}}(\vect{p})\,e^{i\vect{p}\cdot\vect{x}}\,,\qquad
\widetilde\Psi_{i,\vect{k}}(\vect{p}) = \int d^3x\, \Psi_{i,\vect{k}}(\vect{x})\,e^{-i\vect{p}\cdot\vect{x}}
\end{equation}
upon applying the standard dictionary between discrete and continuum expressions:
\begin{equation}
\sum_{\vect{p}} = V\int\frac{d^3p}{(2\pi)^3} \,,\qquad
\delta_{\vect{p}_1,\vect{p}_2} = \frac{(2\pi)^3}{V}\,\delta^3(\vect{p}_1-\vect{p}_2)\,.
\end{equation}

We now examine the matrix element $\langle i',\vect{k}'| \,\mathcal{O}_1\,e^{i\vect{q}\cdot\vect{x}} \,| i, \vect{k}\rangle$ involved in Eq.~\eqref{eq:diagram1} for the $\bar v^2$ and $\mathbb{1}$ operators; $\langle i,\vect{k}| \,\mathcal{O}_2\,e^{-i\vect{q}\cdot\vect{x}} \,| i', \vect{k}'\rangle$ is completely analogous. 
For the $\bar v^2$ operator, we simply obtain
\begin{align}
\langle i',\vect{k}'| \,\bar v^2\,e^{i\vect{q}\cdot\vect{x}} \,| i, \vect{k}\rangle 
=&\; -\frac{1}{8m_e^2} \int d^3x \,\bigl(\Psi_{i',\vect{k}'}^* \overleftrightarrow{\nabla}^2\, \Psi_{i,\vect{k}} \bigr)\,e^{i\vect{q}\cdot\vect{x}} \nonumber\\
=&\; \frac{1}{8m_e^2} \sum_{\vect{G}',\vect{G}} (\vect{k}'+\vect{G}' +\vect{k}+\vect{G})^2\, u^*_{i',\vect{k}',\vect{G}'} \, u_{i,\vect{k},\vect{G}} \, \delta_{\vect{k}'+\vect{G}',\,\vect{k}+\vect{G}+\vect{q}} \,.
\label{eq:expectation_v2_semi}
\end{align}
For NR absorption in the mass range of interest here, $m_\phi \lesssim 100$ eV, the momentum transfer $q\sim 10^{-3} m_\phi \sim \text{meV}\, \bigl(\frac{m_\phi}{\text{eV}}\bigr)$ is well within the 1BZ ($\mathcal{O}$(keV)). 
This implies that Umklapp processes where $\vect{G}'\ne\vect{G}$ do not contribute, so (lattice) momentum conservation simply dictates $\vect{k}'=\vect{k}+\vect{q}$. 
At leading order in $q$ we can set $\vect{k}' = \vect{k}$, and Eq.~\eqref{eq:expectation_v2_semi} simplifies to
\begin{equation}
\langle i',\vect{k}'| \,\bar v^2\,e^{i\vect{q}\cdot\vect{x}} \,| i, \vect{k}\rangle 
=\delta_{\vect{k}',\vect{k}}\,\frac{1}{2m_e^2} \sum_{\vect{G}} (\vect{k}+\vect{G})^2 \, u^*_{i', \vect{k}, \vect{G}} \,u_{i, \vect{k},\vect{G}} +\mathcal{O}(q) \,.
\end{equation}
For the $\mathbb{1}$ operator, additional care is needed since $\langle i',\vect{k}'| \,e^{i\vect{q}\cdot\vect{x}} \,| i, \vect{k}\rangle$ vanishes in the $q\to 0$ limit: $|i',\vect{k}'\rangle$ and $|i,\vect{k}\rangle$ are distinct energy eigenstates and therefore orthogonal.
At $\mathcal{O}(q)$, we have $\langle i',\vect{k}'| \,e^{i\vect{q}\cdot\vect{x}} \,| i, \vect{k}\rangle\simeq i\vect{q}\cdot \langle i',\vect{k}'| \,\vect{x} \,| i, \vect{k}\rangle$. 
A numerically efficient way to compute this matrix element is to trade the position operator for the momentum operator via its commutator with the Hamiltonian $H=\frac{\vect{p}^2}{2m_e} + V(\vect{x})$:
\begin{equation}
\langle i',\vect{k}'| \,\vect{x} \,| i, \vect{k}\rangle 
= -\frac{1}{\varepsilon_{i',\vect{k}'} -\varepsilon_{i,\vect{k}}} \langle i',\vect{k}'| \,[\vect{x}, H] \,| i, \vect{k}\rangle 
= -\frac{i}{m_e(\varepsilon_{i',\vect{k}'} -\varepsilon_{i,\vect{k}})} \langle i',\vect{k}'| \,\vect{p} \,| i, \vect{k}\rangle \,.
\label{eq:trick}
\end{equation}
Substituting in the wave functions, we find:
\begin{equation}
\langle i',\vect{k}'| \,e^{i\vect{q}\cdot\vect{x}} \,| i, \vect{k}\rangle 
=\delta_{\vect{k}',\vect{k}} \,\frac{\vect{q}}{m_e\, \omega_{i'i,\,\vect{k}}} \cdot \sum_{\vect{G}}  (\vect{k}+\vect{G}) \,u^*_{i', \vect{k}, \vect{G}} \,u_{i, \vect{k},\vect{G}} +\mathcal{O}(q^2)\,.
\end{equation}
where $\omega_{i'i,\,\vect{k}}\equiv \varepsilon_{i',\vect{k}} - \varepsilon_{i,\vect{k}}$.

It is convenient to define the following crystal form factors, via which the Bloch wave functions enter DM absorption rates (at leading order in $q$):
\begin{align}
f_{\,i'i,\,\vect{k}} \equiv &\; \frac{1}{2m_e^2} \sum_{\vect{G}} (\vect{k}+\vect{G})^2 \, u^*_{i', \vect{k}, \vect{G}} \,u_{i, \vect{k},\vect{G}}\,, \\
\vect{f}_{i'i,\,\vect{k}} \equiv &\; \frac{1}{\omega_{i'i,\,\vect{k}}}\sum_{\vect{G}}  (\vect{k}+\vect{G}) \,u^*_{i', \vect{k}, \vect{G}} \,u_{i, \vect{k},\vect{G}} \,.
\end{align}
Note that they differ from the crystal form factor used in spin-independent DM scattering~\cite{Essig:2015cda, Trickle:2019nya, Griffin:2021znd}: $f_{[i'\vect{k}',i\vect{k},\vect{G}]}= \sum_{\vect{G}'} u^*_{i',\vect{k}',\vect{G}'+\vect{G}}\, u_{i,\vect{k},\vect{G}'}$. 
The absorption kinematics simply set the $\vect{k}$ and $\vect{G}$ vectors of the initial and final states to be the same; also, powers of $(\vect{k}+\vect{G})$ appear as follows from the effective operators. 

The crystal form factors defined above allow us to write the self-energies in a concise form. 
For the operators $\mathbb{1}$ and $\bar v^2$, the spin trace is trivial and simply yields a factor of two. 
Each pair of valence/conduction states between which a transition can happen contributes to two terms in the sum over electronic states, because either $i,\vect{k}$ or $i',\vect{k}'$ can be a valence or conduction state. 
Combining the two terms for each pair, we obtain
\begin{align}
\Pi_{\mathbb{1},\mathbb{1}} =&\; \frac{2}{V} \sum_{\substack{i'\in\,\text{cond.}\\ i\,\in\,\text{val.}}} \sum_{\vect{k}\,\in\,\text{1BZ}} \left( \frac{1}{\omega -\omega_{i'i,\,\vect{k}}+i\delta} -\frac{1}{\omega +\omega_{i'i,\,\vect{k}}-i\delta}\right) \left| \frac{\vect{q}}{m_e}\cdot\vect{f}_{i'i,\,\vect{k}}\right|^2\,,\label{eq:Pi_11} \\
\Pi_{\bar v^2, \bar v^2} =&\; \frac{2}{V} \sum_{\substack{i'\in\,\text{cond.}\\ i\,\in\,\text{val.}}} \sum_{\vect{k}\,\in\,\text{1BZ}} \left( \frac{1}{\omega -\omega_{i'i,\,\vect{k}}+i\delta} -\frac{1}{\omega +\omega_{i'i,\,\vect{k}}-i\delta}\right) \bigl| f_{i'i,\,\vect{k}}\bigr|^2 \,,\label{eq:Pi_v2v2}
\end{align}
where $\delta \to 0^+$. 
We see explicitly that $\Pi_{\mathbb{1},\mathbb{1}}\sim q^2$ and $\Pi_{\bar v^2, \bar v^2}\sim q^0$, as already alluded to in Sec.~\ref{sec:in-medium}.
The other two self-energies, $\Pi_{\bar v^2, \mathbb{1}}$ and $\Pi_{\mathbb{1}, \bar v^2}$, take the form of $\vect{q}\cdot\vect{\mathcal{F}} +\mathcal{O}(q^2)$, where $\vect{\mathcal{F}}$ is a target-dependent function that involves $f_{\,i'i,\,\vect{k}}$ and $\vect{f}_{\,i'i,\,\vect{k}}$. 
In the absence of a special direction, we must have $\vect{\mathcal{F}} = \vect{0}$ and therefore, $\Pi_{\bar v^2, \mathbb{1}}\,,\, \Pi_{\mathbb{1}, \bar v^2} \sim \mathcal{O}(q^2)$. 
Working out the leading $\mathcal{O}(q^2)$ contribution to these self-energies would require the $\mathcal{O}(q^2)$ term in $\langle i',\vect{k}'| \,e^{i\vect{q}\cdot\vect{x}} \,| i, \vect{k}\rangle$, which however does not admit a simple expression in terms of just the momentum operator as in Eq.~\eqref{eq:trick}. 
Nevertheless, $\Pi_{\bar v^2, \mathbb{1}}$ and $\Pi_{\mathbb{1}, \bar v^2}$ only enter the absorption rate in the scalar DM case and we expect $\mathcal{R}_{\bar v^2, \mathbb{1}} \sim \mathcal{R}_{\mathbb{1}, \mathbb{1}}$ since $\Pi_{\bar v^2, \mathbb{1}}$, $\Pi_{\mathbb{1}, \bar v^2}$ and $\Pi_{\mathbb{1}, \mathbb{1}}$ all scale as $q^2$. 
So as long as $\mathcal{R}_{\mathbb{1}, \mathbb{1}} \ll \mathcal{R}_{\bar v^2, \bar v^2}$, it is justified to neglect the second term in Eq.~\eqref{eq:R_s} altogether and use Eq.~\eqref{eq:R_s_approx} for the rate; computing $\Pi_{\bar v^2, \mathbb{1}}$, $\Pi_{\mathbb{1}, \bar v^2}$ then becomes unnecessary. 
We see from Fig.~\ref{fig:scalar_rate} that this is indeed the case for Si and Ge.

To calculate the DM absorption rates and make sensitivity projections, we use DFT-computed electronic band structures and wave functions for Si and Ge~\cite{sinead_m_griffin_2021_4735777}, including all-electron reconstruction up to a cutoff of 2\,keV; see Ref.~\cite{Griffin:2021znd} for details. 
We adopt the same numerical setup as the ``valence to conduction'' calculation in Ref.~\cite{Griffin:2021znd}, and include also the 3d states in Ge as valence (treating them as core states gives similar results). 
The finite resolution of the $\vect{k}$-grid means we need to apply some kind of smearing to the delta functions coming from the imaginary part of Eqs.~\eqref{eq:Pi_11} and \eqref{eq:Pi_v2v2}. 
This is done in practice by setting $\delta$ in Eqs.~\eqref{eq:Pi_11} and \eqref{eq:Pi_v2v2} to a finite constant 0.2\,eV, which we find appropriate for a $10\times10\times10$ $\vect{k}$-grid for the majority of the DM mass range.
We implement our numerical calculation as a new module ``absorption'' of the \texttt{EXCEED-DM} program~\cite{exceed_dm_collaboration_2021_5009167}.

We present the projected reach for the three DM models in Figs.~\ref{fig:opt_compare_reach} and~\ref{fig:scalar_reach}, assuming 3 events (corresponding to 95\% CL) for 1~kg-yr exposure without including background, together with existing constraints on these models for reference. 
The solid curves are our theoretical predictions; they are obtained using the rate formulae Eqs.~\eqref{eq:R_v}, \eqref{eq:R_ps} and \eqref{eq:R_s_approx} for vector, pseudoscalar and scalar DM, respectively, with the self-energies $\Pi_{\mathbb{1},\mathbb{1}}$, $\Pi_{\bar v^2,\bar v^2}$ computed numerically for Si and Ge according to Eqs.~\eqref{eq:Pi_11} and \eqref{eq:Pi_v2v2} as explained above. 
For pseudoscalar and scalar DM, the reach curves are essentially the sum of Lorentzians coming from the smearing of delta functions in $\text{Im}\,\Pi_{\mathbb{1},\mathbb{1}}$ and $\text{Im}\,\Pi_{\bar v^2, \bar v^2}$, respectively; there is no screening in these cases. 
For vector DM, in-medium mixing with the photon results in the plasmon peak (dip in the reach curves) between 10 and 20\,eV for both Si and Ge; the rate is screened below the plasmon peak.

For vector and pseudoscalar DM, we can alternatively take the data-driven approach, using Eqs.~\eqref{eq:R_v_sigma} and \eqref{eq:R_ps_sigma}, respectively, to derive the rate from the measured conductivity/dielectric. 
As in Ref.~\cite{Hochberg:2016sqx,Bloch:2016sjj}, we use the measured optical data from Ref.~\cite{1985a}. 
Results from this data-driven approach are shown by the dashed curves; they are the same as in Ref.~\cite{Hochberg:2016sqx,Bloch:2016sjj} upon inclusion of backgrounds. 
For Si, the solid and dashed curves are very close to each other for $m_\phi\gtrsim 3$\,eV; the theoretical calculation (solid curves) systematically overestimates the rate as $m_\phi$ approaches the band gap (1.2\,eV) because of the smearing procedure discussed above. 
For Ge, we see the same systematic discrepancy close to the band gap (0.67\,eV); also, the theoretical calculation predicts a sharper plasmon peak (corresponding to a smaller $\text{Im}\,\Pi_{\mathbb{1},\mathbb{1}}$ near the plasmon frequency) compared to data. 
Aside from these issues, we view the overall good agreement between the solid and dashed curves in the vector and pseudoscalar cases as a validation of our DFT-based theoretical calculation in the majority of DM mass range. 
Importantly, this gives credence to the reach curves we have calculated in the scalar DM case, where the data-driven approach does not apply, though one has to keep in mind that our calculation systematically overestimates the rate for DM masses below about 3\,eV because of the smearing issue.

\section{Dark Matter Absorption in Superconductors}
\label{sec:superconductor}

We now turn to the case of conventional superconductors described by BCS theory. 
For the majority of the calculation, we are concerned with electronic states with energies $\varepsilon$ satisfying $|\varepsilon-\varepsilon_F|\gg \Delta$, where $\varepsilon_F$ is the Fermi energy and $2\Delta\sim\mathcal{O}(\text{meV})$ is the gap, and the description of a superconductor approaches that of a normal metal; corrections due to Cooper pairing only become relevant within $\mathcal{O}(\Delta)$ of the Fermi surface. 

Following Refs.~\cite{Hochberg:2015pha,Hochberg:2015fth,Hochberg:2016ajh}, we model the electrons near the Fermi surface with a free-electron dispersion $\varepsilon_{\vect{k}}=\frac{k^2}{m_*}$ and wave function $\Psi_{\vect{k}}(\vect{x}) = \frac{1}{\sqrt{V}}\,e^{i\vect{k}\cdot\vect{x}}$, where the effective mass $m_*$ is generally an $\mathcal{O}(1)$ number times the electron's vacuum mass $m_e$. 
At zero temperature, electrons occupy states up to the Fermi surface, a sphere of radius $k_F^{}=\sqrt{2m_*^{}\varepsilon_F^{}}$. 
The volume of the Fermi sphere gives the density of free electrons, $n_e=\frac{2}{(2\pi)^3}\frac{4}{3}\pi k_F^3$, where the twofold spin degeneracy has been taken into account. 
We expect this simple effective description to hold up to a UV cutoff $\omega_\text{max}$ ($\sim 0.5\,$ eV for Al), above which interband transitions become important and one may instead perform a DFT calculation (as in the case of crystals discussed in Sec.~\ref{sec:semiconductor}).

Within this simple free-electron model, the self-energies $\Pi_{\mathcal{O}_1,\mathcal{O}_2}$ are real at one-loop level; it is easy to see that two electronic states differing by energy $\omega$ and momentum $\vect{q}$ cannot be both on-shell when $\omega\gg q$. 
Therefore, the leading contribution to the imaginary part arises at two loops, and we have
\begin{equation}
\Pi_{\mathcal{O}_1,\mathcal{O}_2} \simeq \text{Re}\,\Pi_{\mathcal{O}_1,\mathcal{O}_2}^\text{(1-loop)} +i\,\text{Im}\,\Pi_{\mathcal{O}_1,\mathcal{O}_2}^\text{(2-loop)} \,.
\end{equation}

For the real part $\text{Re}\,\Pi_{\mathcal{O}_1,\mathcal{O}_2}=\text{Re}\,\Pi_{\mathcal{O}_1,\mathcal{O}_2}^\text{(1-loop)}$, we apply the general formula Eq.~\eqref{eq:diagram1} to the free-electron model in the limit $\omega\gg q$, as explained in detail in App.~\ref{app:self-energy-metal-1}. 
The results are:
\begin{equation}
\text{Re}\,\Pi_{\mathbb{1},\mathbb{1}} = \frac{q^2}{\omega^2}\, \frac{n_e}{m_*} \,,\qquad
\text{Re}\,\Pi_{\bar v^2,\mathbb{1}} = 
\text{Re}\,\Pi_{\mathbb{1},\bar v^2} = \frac{k_F^2}{2m_e^2}\,\frac{q^2}{\omega^2}\,\frac{n_e}{m_*} \,.
\end{equation}
While these are derived for normal conductors, we expect them to carry over to the superconductor case; proportionality to $n_e$ (the total number of electronic states within the Fermi sphere) implies insensitivity to deformations within $\mathcal{O}(\Delta)$ of the Fermi surface.
We also note in passing that, via Eq.~\eqref{eq:dielectric}, we obtain the familiar result for the photon self-energies~\cite{Raffelt:1996wa,Redondo:2013lna}: $\text{Re}\,\Pi_T = \omega_p^2$, $\text{Re}\,\Pi_L = \frac{Q^2}{\omega^2}\,\omega_p^2$, where $\omega_p^2 \equiv \frac{e^2n_e}{m_*}$ is the plasma frequency squared.

For the imaginary part $\text{Im}\,\Pi_{\mathcal{O}_1,\mathcal{O}_2}=\text{Im}\,\Pi_{\mathcal{O}_1,\mathcal{O}_2}^\text{(2-loop)}$, we expect the dominant contribution to come from two-loop diagrams with an internal phonon line for a high-purity sample (otherwise impurity scattering may also contribute). 
These are associated with $\phi$ (or $\gamma$) $+\; e^- \to e^- \;+$ phonon processes by the optical theorem, and can be computed by the standard cutting rules, as we detail in App.~\ref{app:self-energy-metal-2}. 
We model the (acoustic) phonons with a linear dispersion, $\omega_{\vect{q}'} = c_s q'$ where $c_s$ is the sound speed, and neglect Umklapp processes which amounts to imposing a cutoff on the phonon momentum, $q'_\text{max} = q_D^{} \equiv \omega_D^{}/c_s$ with $\omega_D$ the Debye frequency. 
The electron-phonon coupling, in our normalization convention, is given by $\frac{C_\text{e-ph} q'}{\sqrt{2\omega_{\vect{q}'} \rho_T^{}}}$~\cite{Hochberg:2016ajh,Kittel,Mahan}, with $C_\text{e-ph}\sim\mathcal{O}(\varepsilon_F^{})$ a constant with mass dimension one. 
Accounting for the superconducting gap, we obtain, for $\omega\gg q$:
\begin{align}
\text{Im}\,\Pi_{\mathbb{1},\mathbb{1}} =&\; -\frac{C_\text{e-ph}^2\, \omega^2\,q^2}{3\,(2\pi)^3 \rho_T^{}\,c_s^6} \int_0^{\min\left(1-\frac{2\Delta}{\omega},\, \frac{\omega_D^{}}{\omega}\right)} dx\, x^4 (1-x)\, E\left(\sqrt{1-\frac{(2\Delta/\omega)^2}{(1-x)^2}}\,\right) ,
\\
\text{Im}\,\Pi_{\bar v^2, \bar v^2} =&\; -\frac{C_\text{e-ph}^2\, \omega^4}{(2\pi)^3 \rho_T^{}\,c_s^4}\,\frac{m_*^4}{m_e^4} \int_0^{\min\left(1-\frac{2\Delta}{\omega},\, \frac{\omega_D^{}}{\omega}\right)} dx\, x^2 (1-x)^3\, E\left(\sqrt{1-\frac{(2\Delta/\omega)^2}{(1-x)^2}}\,\right) ,
\\
\text{Im}\,\Pi_{\bar v^2,\mathbb{1}} =\text{Im}\,\Pi_{\mathbb{1},\bar v^2} =&\; \frac{\,C_\text{e-ph}^2\, \omega^2\,q^2}{3\,(2\pi)^3 \rho_T^{}\,c_s^4}\,\frac{m_*^2}{m_e^2} \int_0^{\min\left(1-\frac{2\Delta}{\omega},\, \frac{\omega_D^{}}{\omega}\right)} dx\, x^2 (1-x)^2\, E\left(\sqrt{1-\frac{(2\Delta/\omega)^2}{(1-x)^2}}\,\right) ,
\end{align}
where $E(z) = \int_0^1 dt \,\sqrt{\frac{1-z^2t^2}{1-t^2}}$ is the complete elliptic integral of the second kind. 
For energy depositions much higher than the gap, $\omega\gg 2\Delta$, the elliptic integral $E(1)=1$ drops out and we reproduce the results for a normal conductor; see App.~\ref{app:self-energy-metal-2} for details.

%%%
\begin{table}
	\begin{tabular}{ll}
		\hline
		Fermi energy & $\varepsilon_F^{}=11.7$\,eV \\
		Plasma frequency & $\omega_p = 12.2$\,eV \\
		Electron effective mass & $m_*= \frac{9\pi^2\omega_p^4}{128\alpha^2\varepsilon_F^3} = 0.35 \, m_e$ \\
		Fermi momentum & $k_F^{}=\sqrt{2m_*^{}\varepsilon_F^{}} = 2.1$\,keV \\
		Superconducting gap & $2\Delta = 0.6$\,meV \\
		Debye frequency & $\omega_D^{} = 37$\,meV\\
		Sound speed & $c_s = 2.1\times 10^{-5}$ \\
		Maximum phonon momentum\hspace{40pt} & $q_D^{} = \frac{\omega_D^{}}{c_s} = 1.8$\,keV \\
		Electron-phonon coupling & $C_\text{e-ph} = 56$\,eV \\
		Mass density & $\rho_T^{} =2.7$\,g/cm$^3$ \\
		\hline
	\end{tabular}
\caption{\label{tab:Al-params}
	Material parameters for aluminum superconductor.}
\end{table}

With the expressions of self-energies above, we can use Eqs.~\eqref{eq:R_v}, \eqref{eq:R_ps} and \eqref{eq:R_s} to calculate the absorption rates for vector, pseudoscalar and scalar DM. 
We consider an aluminum superconductor (Al-SC) target, for which the relevant material parameters are listed in Table~\ref{tab:Al-params}. 
We use the same numerical values as in Ref.~\cite{Hochberg:2016ajh} for $\varepsilon_F^{}$, $\omega_p$, $\Delta$, $\omega_D^{}$, $c_s$, and determine the electron-phonon coupling $C_\text{e-ph}$ from resistivity measurements~\cite{PhysRevB.3.305,PhysRevB.36.2920} as explained in App.~\ref{app:self-energy-metal-2}. 
For scalar DM, we again confirm the dominance of the $\mathcal{R}_{\bar v^2 \bar v^2}$ term in Eq.~\eqref{eq:R_split}, as seen in Fig.~\ref{fig:scalar_rate}, so the rate formula Eq.~\eqref{eq:R_s} simplifies to Eq.~\eqref{eq:R_s_approx} as in the cases of Si and Ge discussed in Sec.~\ref{sec:semiconductor}.

Figs.~\ref{fig:opt_compare_reach} and~\ref{fig:scalar_reach} show the projected reach, assuming 3 events per kg-yr exposure without including background. 
We see that Al-SC, with its $\mathcal{O}(\text{meV})$ gap, significantly extends the reach with respect to Si and Ge to lower $m_\phi$.
The solid red curves are obtained from the self-energy calculations discussed above; the underlying model has a UV cutoff $\omega_\text{max}\sim0.5$\,eV where we truncate the curves.
Low-temperature conductivity data are available between 0.2\,eV and 3\,eV~\cite{PhysRevB.12.5615}. 
For the vector and pseudoscalar DM models, we also present the reach following the data-driven approach in this mass range (dashed curves), obtained by using Eqs.~\eqref{eq:R_v_sigma} and \eqref{eq:R_ps_sigma} with $\sigma_1$($=\text{Re}\,\sigma=\omega\,\text{Im}\,\varepsilon$) taken from Ref.~\cite{PhysRevB.12.5615} and $\text{Re}\,\varepsilon$ set to $1-\frac{\omega_p^2}{\omega^2}$. 
Between 0.2\,eV and 0.5\,eV where both theoretical (solid) and data-driven (dashed) predictions are shown, they are in reasonable agreement, with the latter stronger by about 40\% for both $\kappa$ and $g_{aee}$ at 0.2\,eV. 
The difference is presumably a result of approaching the UV cutoff of the theoretical calculation, and possibly also the neglect of Umklapp contributions. 
For scalar DM, the data-driven approach is not viable, and we present our theoretical prediction up to 0.5\,eV. 
We also show the reach curves obtained in the previous literature~\cite{Hochberg:2016ajh,Gelmini:2020xir} for comparison, and discuss the differences in what follows.

\paragraph*{\underline{Comparison with previous calculations.}}
The calculation of DM absorption in superconductors was first carried out in Ref.~\cite{Hochberg:2016ajh}, where the 2$\to$2 matrix element for $\phi\, +\, e^- \to e^- +\, \text{phonon}$ was evaluated at leading order in $q$. 
For vector and pseudoscalar DM, our results agree with Ref.~\cite{Hochberg:2016ajh} as seen in Fig.~\ref{fig:opt_compare_reach}, up to a minor numerical prefactor understood as follows.  Ref.~\cite{Hochberg:2016ajh} chose the value of the electron-phonon coupling $C_\text{e-ph}$ such that the photon absorption rate (\ie\ conductivity $\sigma_1$) matches the experimentally measured value at $\omega= 0.2\,$eV. 
In this work, we instead determine $C_\text{e-ph}$ via the $\lambda_\text{tr}$ parameter following Refs.~\cite{PhysRevB.3.305,PhysRevB.36.2920}, which results in a slightly lower value and hence the slight mismatch observed in Fig.~\ref{fig:opt_compare_reach}. 

The more significant numerical difference in the scalar case between our results and Ref.~\cite{Hochberg:2016ajh}, as seen in Fig.~\ref{fig:scalar_reach}, can be traced to two sources.  
First, the numerically dominant effect is that Ref.~\cite{Hochberg:2016ajh} did not distinguish $m_*$ and $m_e$, while we have kept the vacuum mass $m_e$ in the operator coefficients and used the effective mass $m_*$ for the electron's dispersion and phase space; the two masses differ by about a factor of three in Al-SC. 
Note that the difference between $m_e$ and $m_*$ does not affect the vector and pseudoscalar absorption rates as they only depend on $\Pi_{\mathbb{1},\mathbb{1}}$, which is independent of $m_*/m_e$. 
Second, Ref.~\cite{Hochberg:2016ajh} dropped a factor of $(1-x)^2$ in the scalar absorption matrix element when taking the soft phonon limit; this results in an $\mathcal{O}(1)$ difference on the projected reach that is numerically subdominant. 
One can easily verify these two points by evaluating the integral in Eq.~\eqref{eq:ImPi_final} using $x^2 (1-x)$ in place of $\frac{m_*^4}{m_e^4}\,x^2 (1-x)^3$ in the last line; this would reproduce the analytic relation presented in Ref.~\cite{Hochberg:2016ajh} between scalar and photon absorption rates in the limit $\omega\gg 2\Delta$.

More recently, Ref.~\cite{Gelmini:2020xir} revisited scalar DM absorption and claimed that in-medium effects lead to a significantly weaker reach. 
We reiterate that while in-medium mixing with the photon screens the contribution from the LO operator $\mathbb{1}$, the leading contribution to scalar absorption comes instead from the NLO operator $\bar v^2$ that is not screened. 
In fact, the screening factor in Ref.~\cite{Gelmini:2020xir} was (correctly) derived for the $\mathbb{1}$ operator but inconsistently applied to the dominant contribution coming from the $\bar v^2$ operator as obtained in Ref.~\cite{Hochberg:2016ajh}.
As a result, Ref.~\cite{Gelmini:2020xir} significantly underestimated the reach as we can see from Fig.~\ref{fig:scalar_reach}.

\section{Conclusions}

In this paper we revisited the calculation of electronic excitations induced by absorption of bosonic DM. Specifically, we focused on $\mathcal{O}(1\,\text{-}\,100)\,$eV mass DM for Si and Ge targets that are in use in current experiments, and sub-eV mass DM that a proposed Al superconductor detector will be sensitive to.
We utilized an NR EFT framework, where couplings between the DM and electron in a relativistic theory are matched onto NR effective operators in a $1/m_e$ expansion. 
We then computed absorption rates from in-medium self-energies, carefully accounting for mixing between the DM and the photon. 
For crystal targets like Si and Ge, we used first-principles calculations of electronic band structures and wave functions based on density functional theory, and implemented the numerical rate calculation as a new module ``absorption'' of the \texttt{EXCEED-DM} program~\cite{Griffin:2021znd,exceed_dm_collaboration_2021_5009167}. 
For BCS superconductors, we adopted an analytic model as in Refs.~\cite{Hochberg:2015pha,Hochberg:2015fth,Hochberg:2016ajh} treating electrons near the Fermi surface as free quasiparticles and including corrections due to the $\mathcal{O}(\text{meV})$ superconducting gap. 
The projected reach is presented in Figs.~\ref{fig:opt_compare_reach} and \ref{fig:scalar_reach} for vector, pseudoscalar and scalar DM.

Most of previous calculations of DM absorption relied upon relating the process to photon absorption, and hence to the target's optical properties, \ie\ the complex conductivity/dielectric. 
For vector and pseudoscalar DM, this is a valid approach. 
Our theoretical calculations reproduced the results of this data-driven approach in the majority of mass range, which we view as a validation of our methodology and numerical implementation. 

For scalar DM, however, we showed that the dominant contribution is not directly related to photon absorption. 
One therefore cannot simply rescale optical data to derive the DM absorption rate. 
Importantly, the familiar coincidence between scalar and vector couplings, $\bar\psi \psi\simeq \bar\psi\gamma^0\psi$, holds only at leading order in the NR EFT. 
For non-relativistic scalar DM $\phi$, matrix elements of the leading order operator are severely suppressed by the momentum transfer $q\sim 10^{-3} m_\phi$. 
The dominant contribution comes instead from a different operator that is formally next-to-leading-order in the NR EFT expansion, and does not suffer from in-medium screening. 
We presented reach projections for scalar DM based on our theoretical calculations. 
Notably, for Al superconductor, the reach we found is much more optimistic than the recent estimate in Ref.~\cite{Gelmini:2020xir}.

It is straightforward to extend the calculation presented here to anisotropic targets and materials with spin-dependent electronic wave functions (as can arise from spin-orbit coupling); we will investigate this subject in detail in an upcoming publication. 
Another future direction is to calculate phonon and magnon excitations from DM absorption via in-medium self-energies in a similar EFT framework, refining and extending the calculation in Ref.~\cite{Mitridate:2020kly}. 
Finally, in-medium self-energies are also relevant for DM detection via scattering; one can carry out a calculation similar to what we have done here, but in a different kinematic regime, to include in-medium screening corrections in the study of DM-electron scattering via general EFT interactions~\cite{Catena:2021qsr}.

\acknowledgments

We thank Sin\'ead Griffin and Katherine Inzani for DFT calculations used in this work, and Mengxing Ye for helpful discussions. 
This material is based upon work supported by the U.S.\ Department of Energy, Office of Science, Office of High Energy Physics, under Award No.\ DE-SC0021431, by a Simons Investigator Award (K.Z.) and the Quantum Information Science Enabled Discovery (QuantISED) for High Energy Physics (KA2401032). 
The computations presented here were conducted on the Caltech High Performance Cluster, partially supported by a grant from the Gordon and Betty Moore Foundation. 

\appendix

\section{Self-energy Calculations}
\label{app:self-energy}

\subsection{General Result for the One-loop Self-energy}
\label{app:self-energy-1}

At one-loop level, the self-energies defined in Eqs.~\eqref{eq:graph_topology_1} and \eqref{eq:graph_topology_2} are given by
\begin{equation}
-i\,\Pi_{\mathcal{O}_1,\mathcal{O}_2}(Q) = 
\parbox[c][60pt][c]{90pt}{\centering
	\begin{fmffile}{diags/se1-1loop}
	\begin{fmfgraph*}(70,40)
	\fmfleft{in}
	\fmfright{out}
	\fmf{photon,tension=2,label=$\overset{Q}{\longrightarrow}$,l.side=left,l.d=3pt}{in,v1}
	\fmf{photon,tension=2}{v2,out}
	\fmf{fermion,left,tension=0.5,label={\scriptsize $I'$}}{v1,v2}
	\fmf{fermion,left,tension=0.5,label={\scriptsize $I$}}{v2,v1}
	\fmfv{decor.shape=circle,decor.filled=30,decor.size=3thick,label={\footnotesize $\mathcal{O}_1$\;\;},label.angle=-110,l.d=8pt}{v1}
	\fmfv{decor.shape=circle,decor.filled=30,decor.size=3thick,label={\footnotesize \;\;$\mathcal{O}_2$},label.angle=-70,l.d=8pt}{v2}
	\end{fmfgraph*}
	\end{fmffile}
}
\,,\qquad\quad
-i\,\Pi'_{\mathcal{O}}(Q) = 
\parbox[c][80pt][c]{80pt}{\centering
	\begin{fmffile}{diags/se2-1loop}
	\begin{fmfgraph*}(60,60)
	\fmfleft{in}
	\fmfright{out}
	\fmf{photon,tension=2,label=$\overset{Q}{\longrightarrow}\;\;$,l.side=left,l.d=4pt}{in,v1}
	\fmf{photon,tension=2}{v1,out}
	\fmf{fermion,right,tension=0.6,label={\scriptsize $I$}}{v1,v1}
	\fmfv{decor.shape=circle,decor.filled=30,decor.size=3thick,label={\footnotesize $\mathcal{O}$},label.angle=-90,l.d=8pt}{v1}
	\end{fmfgraph*}
	\end{fmffile}
}
\,,
\label{eq:1loop_diag}
\end{equation}
where the external states (drawn with curly lines for concreteness) can be either spin-0 or spin-1, and the internal electronic states $I, I'$ are summed over. 
Using the in-medium Feynman rules (see \eg\ Ref.~\cite{Mahan}) we obtain, for the first diagram:
\begin{equation}
-i\,\Pi_{\mathcal{O}_1,\mathcal{O}_2} = \frac{(-1)}{V} \sum_{I'I} \int_{-\infty}^\infty\frac{d\varepsilon}{2\pi}\, \frac{\text{tr} \left(\langle I' | \,\mathcal{O}_1\, e^{i\vect{q}\cdot\vect{x}} |I\rangle \langle I | \,\mathcal{O}_2\, e^{-i\vect{q}\cdot\vect{x}} |I'\rangle\right)}{(\varepsilon +\omega -\varepsilon_{I'}^{} + i\delta_{I'}^{})(\varepsilon -\varepsilon_I^{} +i\delta_I^{})} \,,
\label{eq:Pi_exp}
\end{equation}
where $V$ is the total volume, ``tr'' represents the spin trace, and $\delta_{I^{(\prime)}}^{}\equiv \delta \,\text{sgn}(\varepsilon_{I^{(\prime)}}^{}-\varepsilon_F^{})$ with $\delta\to 0^+$. 
Note that the $i\delta$ prescription for electron propagators is different from the vacuum theory, and depends on whether the state is above or below the Fermi energy $\varepsilon_F^{}$; using the correct $i\delta$ prescription is crucial for ensuring causality. 
Meanwhile, the matrix elements coming from the vertices are
\begin{equation}
\langle I' | \,\mathcal{O}_1\, e^{i\vect{q}\cdot\vect{x}} |I\rangle = \int d^3x \,\bigl[\Psi_{I'}^*(\vect{x}) \,\mathcal{O}_1 \Psi_I^{}(\vect{x})\bigr] \,e^{i\vect{q}\cdot\vect{x}} \,,
\end{equation}
and likewise for $\langle I | \,\mathcal{O}_2\, e^{-i\vect{q}\cdot\vect{x}} |I'\rangle$. 
Here $\mathcal{O}_{1,2}$ are matrices in spin space, and may involve spatial derivatives acting on the electronic wave functions. 
For example, for the velocity operator defined in Eq.~\eqref{eq:v_op_def} (which is proportional to the identity matrix in spin space), we have
\begin{equation}
\big\langle I' \big| \,v^j\,e^{i\vect{q}\cdot\vect{x}}\, \big| I\big\rangle 
= -\frac{i}{2m_e}\,\big\langle I' \big| \,\overleftrightarrow{\nabla}_{\hspace{-2pt}j} \,e^{i\vect{q}\cdot\vect{x}}\, \big| I\big\rangle 
=
-\frac{i}{2m_e} \int d^3x\, \bigl[ \Psi_{I'}^* \,(\nabla_j \Psi_I^{}) -(\nabla_j\Psi_{I'}^*)\, \Psi_I^{} \bigr] \,e^{i\vect{q}\cdot\vect{x}}\,.
\end{equation}

We can evaluate the energy integral in Eq.~\eqref{eq:Pi_exp} by examining the pole structure of the integrand in the complex plane. 
If $\delta_{I'}^{}$ and $\delta_I^{}$ have the same sign (\ie\ if both $I'$ and $I$ are above or below the Fermi energy), the two poles are on the same side of the real axis and they have opposite residues; the integral therefore vanishes upon closing the contour via either $+i\infty$ or $-i\infty$. 
So we must have one state above the Fermi energy and one below it, in which case there is one pole on each side of the real axis; closing the contour via either $+i\infty$ or $-i\infty$ to pick up the residue at one of the poles, we obtain
\begin{equation}
\int_{-\infty}^\infty\frac{d\varepsilon}{2\pi}\, \frac{1}{(\varepsilon +\omega -\varepsilon_{I'}^{} + i\delta_{I'}^{})(\varepsilon -\varepsilon_I^{} +i\delta_I^{})} 
=
\begin{cases}
\dfrac{i}{\omega-\omega_{I'I}^{} +i\delta} & \text{if \, $\delta_{I'}^{}>0$, $\delta_I^{}<0$} \,;\\[12pt]
-\dfrac{i}{\omega-\omega_{I'I}^{} -i\delta} & \text{if \, $\delta_{I'}^{}<0$, $\delta_I^{}>0$} \,.
\end{cases}
\end{equation}
Here $\omega_{I'I}^{}\equiv \varepsilon_{I'}^{} -\varepsilon_I^{}$, and $\delta\to0^+$.
All cases discussed above can be concisely summarized as:
\begin{equation}
\int_{-\infty}^\infty\frac{d\varepsilon}{2\pi}\, \frac{1}{(\varepsilon +\omega -\varepsilon_{I'}^{} + i\delta_{I'}^{})(\varepsilon -\varepsilon_I^{} +i\delta_I^{})} 
= \frac{-i\, (f_{I'}^{}-f_I^{})}{\omega-\omega_{I'I}^{} +i\delta_{I'I}^{}} \,,
\end{equation}
where $f_I^{}$, $f_{I'}^{}$ are the occupation numbers (equal to 1 for states below the Fermi energy, 0 for states above it), and $\delta_{I'I}^{}\equiv \delta \,\text{sgn}(\omega_{I'I}^{})$. 
We therefore obtain
\begin{equation}
\Pi_{\mathcal{O}_1,\mathcal{O}_2} =
-\frac{1}{V}\, \sum_{I'I}\, \frac{f_{I'}-f_I}{\omega-\omega_{I'I}+i\delta_{I'I}^{}} 
\,\text{tr} \Bigl(\langle I' | \,\mathcal{O}_1 \,e^{i\vect{q}\cdot\vect{x}} | I\rangle \langle I | \,\mathcal{O}_2 \,e^{-i\vect{q}\cdot\vect{x}} | I'\rangle \Bigr) \,.
\label{eq:diagram1}
\end{equation}
As the simplest example, setting $\mathcal{O}_1=\mathcal{O}_2=\mathbb{1}$ in Eq.~\eqref{eq:diagram1}, we obtain $\Pi_{\mathbb{1},\mathbb{1}}$, and hence the dielectric via Eq.~\eqref{eq:dielectric}, which reproduces the familiar Lindhard formula (see \eg\ Ref.~\cite{Dressel} and recent discussions in Refs.~\cite{Hochberg:2021pkt,Knapen:2021run}). 

Now move on to the second diagram in Eq.~\eqref{eq:1loop_diag}. 
While we have shown in Sec.~\ref{sec:in-medium} that contributions to absorption rates from this diagram can be eliminated using the Ward identity, we present its result here for completeness and also to allow for an explicit check of the Ward identity.
In this diagram, the electron propagator starts and ends at the same time point and time-ordering becomes ambiguous. 
The correct prescription is to take the normal-ordered product of creation and annihilation operators, and the loop is simply proportional to the electron number operator~\cite{Mahan}. 
Again writing the result in terms of occupation number $f_I^{}$, we find
\begin{equation}
\Pi'_{\mathcal{O}} = -\frac{1}{V}\, \sum_{I}\, f_I
\,\text{tr} \langle I | \,\mathcal{O} | I\rangle  \,.
\label{eq:diagram2}
\end{equation}
Note that $\Pi'_{\mathcal{O}}$ is purely real at all orders. 
With Eqs.~\eqref{eq:diagram1} and \eqref{eq:diagram2} one can readily verify the relations implied by the Ward identity, Eqs.~\eqref{eq:relations-1} and \eqref{eq:relations-2}.

\subsection{Real Part of the One-loop Self-energy in a Metal}
\label{app:self-energy-metal-1}

We now apply Eq.~\eqref{eq:diagram1} to the case of a metal.
As discussed in Sec.~\ref{sec:superconductor}, we model the electrons near the Fermi surface of a metal as free quasiparticles with an effective mass $m_*$ and energy eigenstates labeled by momentum.
The sum over $I, I'$ becomes integrals over $\vect{k},\vect{k}'$, and we have
\begin{equation}
\Pi_{\mathcal{O}_1,\mathcal{O}_2} = \text{Re}\,\Pi_{\mathcal{O}_1,\mathcal{O}_2} =
-\frac{1}{V} \int \frac{V\,d^3k'}{(2\pi)^3} \int \frac{V\,d^3k}{(2\pi)^3} \frac{f_{\vect{k}'}-f_{\vect{k}}}{\omega-\frac{k'^2}{2m_*} +\frac{k^2}{2m_*}} 
\,\text{tr} \Bigl(\langle \vect{k}' | \,\mathcal{O}_1 \,e^{i\vect{q}\cdot\vect{x}} | \vect{k}\rangle \langle \vect{k} | \,\mathcal{O}_2 \,e^{-i\vect{q}\cdot\vect{x}} | \vect{k}'\rangle \Bigr) \,.
\end{equation}
Note that the $i\delta$ in the denominator is irrelevant since the intermediate states cannot go on-shell and $\Pi_{\mathcal{O}_1,\mathcal{O}_2}$ is real at one-loop level. 

Let us first consider $\Pi_{\mathbb{1},\mathbb{1}}$. 
For the matrix element part, we have
\begin{align}
\langle \vect{k}' | \,\mathbb{1}\,e^{i\vect{q}\cdot\vect{x}} | \vect{k} \rangle =&\; \frac{1}{V} \int d^3x \,e^{i(\vect{k}+\vect{q}-\vect{k}')\cdot\vect{x}}\,\mathbb{1} = \frac{(2\pi)^3}{V}\,\delta^3(\vect{k}+\vect{q}-\vect{k}') \,\mathbb{1}\,,\\
\text{tr} \Bigl( \langle \vect{k}' | \,\mathbb{1}\,e^{i\vect{q}\cdot\vect{x}} | \vect{k} \rangle \langle \vect{k} | \,\mathbb{1}\,e^{-i\vect{q}\cdot\vect{x}} | \vect{k}' \rangle\Bigr) =&\;  2\,\frac{(2\pi)^3}{V}\,\delta^3(\vect{k}+\vect{q}-\vect{k}') \,\frac{1}{V}\int d^3x
= 2\,\frac{(2\pi)^3}{V}\,\delta^3(\vect{k}+\vect{q}-\vect{k}') \,.
\end{align}
Therefore,
\begin{equation}
\Pi_{\mathbb{1},\mathbb{1}} 
= -2\int\frac{d^3k}{(2\pi)^3} \,\frac{f_{\vect{k}+\vect{q}}-f_{\vect{k}}}{\omega-\frac{(\vect{k}+\vect{q})^2}{2m_*} +\frac{k^2}{2m_*}}  
= -2\int\frac{d^3k}{(2\pi)^3} \,\frac{f_{\vect{k}+\vect{q}}-f_{\vect{k}}}{\omega-\frac{\vect{k}\cdot\vect{q}}{m_*} -\frac{q^2}{2m_*}}  \,.
\end{equation}
Expanding in small $q$ and integrating by parts, we find
\begin{align}
\Pi_{\mathbb{1},\mathbb{1}} 
=&\; -2\int\frac{d^3k}{(2\pi)^3} \left( \vect{q}\cdot\nabla f_{\vect{k}} +\dots\right)  \left(\frac{1}{\omega} +\frac{\vect{k}\cdot\vect{q}}{m_*\omega^2} +\dots\right) \nonumber\\
=&\; 2\int\frac{d^3k}{(2\pi)^3} \,f_{\vect{k}} \left(\vect{q}\cdot\nabla +\dots\right) \left(\frac{1}{\omega} +\frac{\vect{k}\cdot\vect{q}}{m_*\omega^2} +\dots\right) \nonumber\\
\simeq &\; 2\int\frac{d^3k}{(2\pi)^3} \,f_{\vect{k}}\, \frac{q^2}{m_*\omega^2}
\,=\, \frac{q^2}{\omega^2} \,\frac{n_e}{m_*}\,,
\end{align}
where the gradients are in $\vect{k}$ space, and $n_e=2\int\frac{d^3k}{(2\pi)^3} \,f_{\vect{k}}$ is the free electron density.

We can calculate $\Pi_{\bar v^2, \mathbb{1}}$ in a similar way. 
The matrix element part again yields a momentum-conserving delta function, and the integrand can then be expanded in small $q$. 
We find
\begin{align}
\Pi_{\bar v^2,\mathbb{1}} 
=&\; -2\int\frac{d^3k}{(2\pi)^3} \,\frac{f_{\vect{k}+\vect{q}}-f_{\vect{k}}}{\omega-\frac{\vect{k}\cdot\vect{q}}{m_*} -\frac{q^2}{2m_*}} \,\frac{(2\vect{k}+\vect{q})^2}{8m_e^2} \nonumber\\
=&\; 2\int\frac{d^3k}{(2\pi)^3}\, f_{\vect{k}} \left(\vect{q}\cdot\nabla - \frac{1}{2}\, q^i q^j\, \nabla_i \nabla_j+\dots\right)
\left(\frac{1}{\omega} +\frac{\vect{k}\cdot\vect{q}}{m_*\omega^2} +\dots\right)
\left( \frac{k^2}{2m_e^2} +\frac{\vect{k}\cdot\vect{q}}{2m_e^2} +\dots\right)
\nonumber\\
\simeq&\; 2\int\frac{d^3k}{(2\pi)^3}\, f_{\vect{k}} 
\,\frac{k^2 q^2 +2(\vect{k}\cdot\vect{q})^2}{2m_e^2 m_* \omega^2}
\,=\, \frac{k_F^2}{2m_e^2} \,\frac{q^2}{\omega^2}\, \frac{n_e}{m_*} \,,
\label{eq:Re_Pi_v2_1_metal}
\end{align}
where we have used $f_{\vect{k}} = \Theta(k_F^{}-k)$, and $n_e=\frac{2}{(2\pi)^3}\frac{4}{3}\pi k_F^3$. 
Finally, since Eq.~\eqref{eq:Re_Pi_v2_1_metal} is invariant under $(\omega, \vect{q}) \to (-\omega, -\vect{q})$, we have $\Pi_{\mathbb{1}, \bar v^2}(Q)=\Pi_{\bar v^2, \mathbb{1}}(-Q)=\Pi_{\bar v^2, \mathbb{1}}(Q)$.

\subsection{Imaginary Part of the Two-loop Self-energy in a Metal}
\label{app:self-energy-metal-2}

The one-loop self-energies calculated above are purely real: both electrons cannot go on-shell if their energies and momenta differ by $Q^\mu = (\omega, \vect{q})$ with $\omega\gg q$. 
The leading contribution to $\text{Im}\,\Pi_{\mathcal{O}_1,\mathcal{O}_2}(Q)$ comes from two-loop diagrams with an internal phonon line. 
In this section, we compute them first in the case of a normal conductor, and then discuss the corrections needed in the superconductor case when $\omega$ approaches the gap $2\Delta$.

\paragraph*{\underline{Cut diagrams.}}
There are three contributing diagrams:
\begin{align}
\parbox[c][90pt][c]{90pt}{\centering
	\begin{fmffile}{diags/se1-2loop-1}
	\begin{fmfgraph*}(80,80)
	\fmfleft{in}
	\fmfright{out}
	\fmftop{t1,t2}
	\fmfbottom{b1,b2}
	\fmf{photon,tension=1,label=$\overset{Q}{\longrightarrow}$,l.side=left,l.d=3pt}{in,v1}
	\fmf{photon,tension=1}{v2,out}
	\fmf{phantom,tension=1.5}{t1,va}
	\fmf{phantom,tension=1.5}{t2,vb}
	\fmf{phantom,tension=1.5}{b1,vc}
	\fmf{phantom,tension=1.5}{b2,vd}
	\fmf{fermion,left=0.25}{v1,va}
	\fmf{fermion,left=0.35,tension=2}{va,vb}
	\fmf{fermion,left=0.25}{vb,v2}
	\fmf{plain,left=0.25}{v2,vd}
	\fmf{fermion,left=0.35,tension=2,label={\scriptsize $K$}}{vd,vc}
	\fmf{plain,left=0.25}{vc,v1}
	\fmf{dashes,right=0.8,tension=0,label={\;$\underset{Q'}{\rightarrow}$},l.d=3pt}{va,vb}
	\fmfv{decor.shape=circle,decor.filled=30,decor.size=3thick,label={\footnotesize $\mathcal{O}_1$\;\;},label.angle=-110,l.d=8pt}{v1}
	\fmfv{decor.shape=circle,decor.filled=30,decor.size=3thick,label={\footnotesize \;\;\;$\mathcal{O}_2$},label.angle=-70,l.d=8pt}{v2}
	\end{fmfgraph*}
	\end{fmffile}
}
=&\; -i \int\frac{d^4K}{(2\pi)^4} \int\frac{d^4Q'}{(2\pi)^4} \,G_K^{}\,G_{K+Q}^{} \, G_{K+Q-Q'}^{} \,G_{K+Q}^{} \, G_{Q'}^\text{ph} \nonumber\\[-25pt]
&\hspace{110pt} y_{q'}^2 \,\text{tr} \Bigl[\widetilde{\mathcal{O}}_1(K, K+Q) \,\widetilde{\mathcal{O}}_2(K+Q, K)\Bigr] \,, \label{eq:2loop_diag-1} \\
\parbox[c][90pt][c]{90pt}{\centering
	\begin{fmffile}{diags/se1-2loop-2}
	\begin{fmfgraph*}(80,80)
	\fmfleft{in}
	\fmfright{out}
	\fmftop{t1,t2}
	\fmfbottom{b1,b2}
	\fmf{photon,tension=1,label=$\overset{Q}{\longrightarrow}$,l.side=left,l.d=3pt}{in,v1}
	\fmf{photon,tension=1}{v2,out}
	\fmf{phantom,tension=1.5}{t1,va}
	\fmf{phantom,tension=1.5}{t2,vb}
	\fmf{phantom,tension=1.5}{b1,vc}
	\fmf{phantom,tension=1.5}{b2,vd}
	\fmf{plain,left=0.25}{v1,va}
	\fmf{fermion,left=0.35,tension=2,label={\scriptsize $K+Q$},l.d=5pt}{va,vb}
	\fmf{plain,left=0.25}{vb,v2}
	\fmf{fermion,left=0.25}{v2,vd}
	\fmf{fermion,left=0.35,tension=2}{vd,vc}
	\fmf{fermion,left=0.25}{vc,v1}
	\fmf{dashes,right=0.8,tension=0,label={ $\overset{\;Q'}{\rightarrow}$\;},l.d=2pt}{vd,vc}
	\fmfv{decor.shape=circle,decor.filled=30,decor.size=3thick,label={\footnotesize $\mathcal{O}_1$\;\;\;},label.angle=-110,l.d=8pt}{v1}
	\fmfv{decor.shape=circle,decor.filled=30,decor.size=3thick,label={\footnotesize \;\;\;\;$\mathcal{O}_2$},label.angle=-70,l.d=8pt}{v2}
	\end{fmfgraph*}
	\end{fmffile}
}
=&\; -i \int\frac{d^4K}{(2\pi)^4} \int\frac{d^4Q'}{(2\pi)^4} \,G_K^{}\,G_{K+Q}^{} \, G_K^{} \,G_{K+Q'}^{} \, G_{Q'}^\text{ph} \nonumber\\[-25pt]
&\hspace{110pt} y_{q'}^2 \,\text{tr} \Bigl[\widetilde{\mathcal{O}}_1(K, K+Q) \,\widetilde{\mathcal{O}}_2(K+Q, K)\Bigr] \,,\label{eq:2loop_diag-2}\\
\parbox[c][90pt][c]{90pt}{\centering
	\begin{fmffile}{diags/se1-2loop-3}
	\begin{fmfgraph*}(80,80)
	\fmfleft{in}
	\fmfright{out}
	\fmftop{t1}
	\fmfbottom{b1}
	\fmf{photon,tension=2.25,label=$\overset{Q}{\longrightarrow}$,l.side=left,l.d=3pt}{in,v1}
	\fmf{photon,tension=2.25}{v2,out}
	\fmf{phantom,tension=3}{t1,va}
	\fmf{phantom,tension=3}{b1,vb}
	\fmf{fermion,left=0.45}{v1,va}
	\fmf{fermion,left=0.45}{va,v2}
	\fmf{fermion,left=0.45}{v2,vb}
	\fmf{fermion,left=0.45,label={\scriptsize $K$},l.d=4pt}{vb,v1}
	\fmf{dashes,tension=0,label={\scriptsize \;\;$\downarrow \;Q'$},l.d=0pt}{va,vb}
	\fmfv{decor.shape=circle,decor.filled=30,decor.size=3thick,label={\footnotesize $\mathcal{O}_1$\;\;},label.angle=-110,l.d=8pt}{v1}
	\fmfv{decor.shape=circle,decor.filled=30,decor.size=3thick,label={\footnotesize \;\;\;$\mathcal{O}_2$},label.angle=-70,l.d=8pt}{v2}
	\end{fmfgraph*}
	\end{fmffile}
}
=&\; -i \int\frac{d^4K}{(2\pi)^4} \int\frac{d^4Q'}{(2\pi)^4} \,G_K^{}\,G_{K+Q}^{} \, G_{K+Q-Q'}^{} \,G_{K-Q'}^{} \, G_{Q'}^\text{ph} \nonumber\\[-25pt]
&\hspace{110pt} y_{q'}^2 \,\text{tr} \Bigl[\widetilde{\mathcal{O}}_1(K, K+Q) \,\widetilde{\mathcal{O}}_2(K+Q-Q', K-Q')\Bigr] \,.\label{eq:2loop_diag-3}
\end{align}
Here each propagator is labeled by a four-momentum that consists of the energy it carries and the momentum label of the electron or phonon state. 
In each diagram, we denote four-momentum flowing into the $\mathcal{O}_1$ vertex from the electron propagator as $K^\mu=(\varepsilon, \vect{k})$, and the phonon four-momentum (with direction indicated by the arrow) as $Q'^\mu = (\omega', \vect{q}')$. 
The electron and phonon propagators are denoted by $iG$ and $iG^\text{ph}$, respectively, with
\begin{equation}
G_K = \frac{1}{\varepsilon -\frac{k^2}{2m_*} +i\delta_\varepsilon} \,,\qquad
G_{Q'}^\text{ph} = \frac{1}{\omega'- \omega_{q'} +i\delta} -\frac{1}{\omega'+\omega_{q'} -i\delta} = \frac{2\,\omega_{q'}}{\omega'^2 - \omega_{q'}^2 +i\delta} \,,
\end{equation}
where $\delta_\varepsilon = \delta \,\text{sgn}(\varepsilon-\varepsilon_F^{})$, $\delta\to 0^+$, and $\omega_{q'} = c_s q'$. 
The electron-phonon vertex $y_{q'} = \frac{C_\text{e-ph} q'}{\sqrt{2\omega_{q'}\rho_T^{}}}$, while the vertices associated with operator insertions $\mathcal{O}_{1,2}$ yield the momentum space representations of these operators, $\widetilde{\mathcal{O}}_{1,2}$, whose arguments are the incoming and outgoing electrons' four-momenta. 
We have assumed exact momentum conservation and neglected Umklapp processes; the latter may introduce an $\mathcal{O}(1)$ correction to the final results which is more difficult to calculate.

By the optical theorem, $2\,\text{Im}\,\Pi_{\mathcal{O}_1,\mathcal{O}_2}$ is given by the sum of cut diagrams. 
For the first diagram, Eq.~\eqref{eq:2loop_diag-1}, there is only one possible cut to put all intermediate states on-shell, \ie\ the one through the phonon propagator and the two electron propagators carrying momenta $K$ and $K+Q-Q'$. 
By the cutting rules, we should replace 
\begin{align}
G_{Q'}^\text{ph}\to &\; -2\pi i \bigl[\delta(\omega'-\omega_{q'}) +\delta(\omega'+\omega_{q'})\bigr] \,,\\
G_K\to &\; -2\pi i\,\text{sgn}(\varepsilon-\varepsilon_F^{})\,\delta\biggl(\varepsilon-\frac{k^2}{2m_*}\biggr) \,,\\
G_{K+Q-Q'}\to &\; -2\pi i\,\text{sgn}(\varepsilon +\omega -\omega' -\varepsilon_F^{})\,\delta\biggl(\varepsilon+\omega -\omega'-\frac{(\vect{k}+\vect{q}-\vect{q}')^2}{2m_*}\biggr) \,.
\end{align}
For $\omega>0$, the on-shell condition requires $\omega'>0$, $\varepsilon<\varepsilon_F^{}$ and $\varepsilon +\omega -\omega'>\varepsilon_F^{}$; this corresponds to a process where an electron jumps out of the Fermi sphere by absorbing $Q^\mu = (\omega, \vect{q})$ while emitting a phonon to conserve momentum.
We therefore obtain
\begin{align}
\text{Cut}\bigl[\,\text{Eq.~\eqref{eq:2loop_diag-1}}\,\bigr] =&\; -
\int\frac{d^4K}{(2\pi)^3} \int\frac{d^4Q'}{(2\pi)^3} \, 2\pi\,\delta(\omega'-\omega_{q'}) \,\delta\biggl(\varepsilon - \frac{k^2}{2m_*}\biggr) \,\delta\biggl(\varepsilon +\omega -\omega' - \frac{(\vect{k}+\vect{q}-\vect{q}')^2}{2m_*}\biggr)  \nonumber\\
&\hspace{100pt} f_{\vect{k}}\,\bigl(1-f_{\vect{k}+\vect{q}-\vect{q}'}\bigr)\,G_{K+Q}^2 \, y_{q'}^2 \,\text{tr} \Bigl[\widetilde{\mathcal{O}}_1(K, K+Q) \,\widetilde{\mathcal{O}}_2(K+Q, K)\Bigr] \nonumber\\[8pt]
=&\; -\int\frac{d^3k}{(2\pi)^3} \int\frac{d^3q'}{(2\pi)^3} \, 2\pi\,\delta\biggl(\omega +\frac{k^2}{2m_*} - \frac{(\vect{k}+\vect{q}-\vect{q}')^2}{2m_*} -\omega_{q'}\biggr) \,f_{\vect{k}}\,\bigl(1-f_{\vect{k}+\vect{q}-\vect{q}'}\bigr) \,y_{q'}^2 \nonumber\\
&\hspace{100pt} G_{K+Q}^2 \,\text{tr} \Bigl[\widetilde{\mathcal{O}}_1(K, K+Q) \,\widetilde{\mathcal{O}}_2(K+Q, K)\Bigr] \,,
\label{eq:cut1}
\end{align}
where it is understood that $\varepsilon$ (the energy components of $K$) is set to $\frac{k^2}{2m_*}$ in the final expression. 
The second diagram, Eq.~\eqref{eq:2loop_diag-2}, is completely analogous. 
Cutting the propagators $G_{K+Q}$, $G_{Q'}^\text{ph}$ and $G_{K+Q'}$, we obtain
\begin{align}
\text{Cut}\bigl[\,\text{Eq.~\eqref{eq:2loop_diag-2}}\,\bigr] =&\; -\int\frac{d^3k}{(2\pi)^3} \int\frac{d^3q'}{(2\pi)^3} \, 2\pi\,\delta\biggl(\omega +\frac{(\vect{k}+\vect{q}')^2}{2m_*} - \frac{(\vect{k}+\vect{q})^2}{2m_*} -\omega_{q'}\biggr) \,f_{\vect{k}+\vect{q}'}\,\bigl(1-f_{\vect{k}+\vect{q}}\bigr) \, y_{q'}^2 \nonumber\\[4pt]
&\hspace{100pt} G_K^2 \,\text{tr} \Bigl[\widetilde{\mathcal{O}}_1(K, K+Q) \,\widetilde{\mathcal{O}}_2(K+Q, K) \Bigr]\nonumber\\[8pt]
=&\; -\int\frac{d^3k}{(2\pi)^3} \int\frac{d^3q'}{(2\pi)^3} \, 2\pi\,\delta\biggl(\omega +\frac{k^2}{2m_*} - \frac{(\vect{k}+\vect{q}-\vect{q}')^2}{2m_*} -\omega_{q'}\biggr)\, f_{\vect{k}}\,\bigl(1-f_{\vect{k}+\vect{q}-\vect{q}'}\bigr)\, y_{q'}^2 \nonumber\\[4pt]
&\hspace{55pt} 
G_{K-Q'}^2  \,\text{tr} \Bigl[\widetilde{\mathcal{O}}_1(K-Q', K+Q-Q') \,\widetilde{\mathcal{O}}_2(K+Q-Q', K-Q')\Bigr] \,,
\label{eq:cut2}
\end{align}
where we have shifted the integration variable $\vect{k}\to\vect{k}-\vect{q}'$ to arrive at the last line.

For the last diagram, Eq.~\eqref{eq:2loop_diag-3}, there are two possible cuts: through $G_K$, $G_{Q'}^\text{ph}$, $G_{K+Q-Q'}$ and through $G_{K+Q}$, $G_{Q'}^\text{ph}$, $G_{K-Q'}$. 
Carrying out the same procedure as above, we obtain
\begin{align}
\text{Cut}\bigl[\,\text{Eq.~\eqref{eq:2loop_diag-3}}\,\bigr] =&\;
-\int\frac{d^3k}{(2\pi)^3} \int\frac{d^3q'}{(2\pi)^3} \,\biggl\{
2\pi\,\delta\biggl(\omega +\frac{k^2}{2m_*} - \frac{(\vect{k}+\vect{q}-\vect{q}')^2}{2m_*} -\omega_{q'}\biggr) \,f_{\vect{k}}\,\bigl(1-f_{\vect{k}+\vect{q}-\vect{q}'}\bigr) \,y_{q'}^2
 \nonumber\\[4pt]
&\hspace{100pt}  G_{K+Q}^{} \,G_{K-Q'}^{} \,\text{tr} \Bigl[\widetilde{\mathcal{O}}_1(K, K+Q) \,\widetilde{\mathcal{O}}_2(K+Q-Q', K-Q') \Bigr]
\nonumber\\[6pt]
&\hspace{90pt} 
+2\pi\,\delta\biggl(\omega +\frac{(\vect{k}-\vect{q}')^2}{2m_*} - \frac{(\vect{k}+\vect{q})^2}{2m_*} -\omega_{q'}\biggr) \,f_{\vect{k}-\vect{q}'}\,\bigl(1-f_{\vect{k}+\vect{q}}\bigr) \, y_{q'}^2
\nonumber\\[4pt]
&\hspace{100pt}
G_{K}^{} \,G_{K+Q-Q'}^{} \,\text{tr} \Bigl[\widetilde{\mathcal{O}}_1(K, K+Q) \,\widetilde{\mathcal{O}}_2(K+Q-Q', K-Q')\Bigr] \biggr\} 
\nonumber\\[8pt]
=&\; 
-\int\frac{d^3k}{(2\pi)^3} \int\frac{d^3q'}{(2\pi)^3} \,
2\pi\,\delta\biggl(\omega +\frac{k^2}{2m_*} - \frac{(\vect{k}+\vect{q}-\vect{q}')^2}{2m_*} -\omega_{q'}\biggr) \,f_{\vect{k}}\,\bigl(1-f_{\vect{k}+\vect{q}-\vect{q}'}\bigr) \,y_{q'}^2
 \nonumber\\[6pt]
&\hspace{70pt}  G_{K+Q}^{} \,G_{K-Q'}^{} \,\text{tr}\Bigl[\widetilde{\mathcal{O}}_1(K, K+Q) \,\widetilde{\mathcal{O}}_2(K+Q-Q', K-Q') 
\nonumber\\[4pt]
&\hspace{140pt} +\widetilde{\mathcal{O}}_1(K-Q', K+Q-Q') \,\widetilde{\mathcal{O}}_2(K+Q, K) \Bigr] \,,
\label{eq:cut3}
\end{align}
where we have shifted the integration variable $\vect{k}\to\vect{k}+\vect{q}'$ and then changed $\vect{q}\to-\vect{q}'$ (assuming the phonon energies $\omega_{q'}$ and electron-phonon couplings $y_{q'}$ depend only on the magnitude but not the direction of $\vect{q}'$) in the second term.

Adding up Eqs.~\eqref{eq:cut1}, \eqref{eq:cut2} and \eqref{eq:cut3}, we obtain
\begin{align}
2\,\text{Im}\,\Pi_{\mathcal{O}_1,\mathcal{O}_2} =&\;
-\int\frac{d^3k}{(2\pi)^3} \int\frac{d^3q'}{(2\pi)^3} \, 
2\pi\,\delta\biggl(\omega +\frac{k^2}{2m_*} - \frac{(\vect{k}+\vect{q}-\vect{q}')^2}{2m_*} -\omega_{q'}\biggr) \,f_{\vect{k}}\,\bigl(1-f_{\vect{k}+\vect{q}-\vect{q}'}\bigr) \,y_{q'}^2
\nonumber\\[4pt]
&\hspace{55pt} \text{tr} \biggl\{\Bigl[ G_{K+Q}\,\widetilde{\mathcal{O}}_1(K, K+Q) + G_{K-Q'} \,\widetilde{\mathcal{O}}_1(K-Q', K+Q-Q')\Bigr]
\nonumber\\[4pt]
&\hspace{70pt} \times\Bigl[ G_{K+Q}\,\widetilde{\mathcal{O}}_2(K+Q, K) + G_{K-Q'} \,\widetilde{\mathcal{O}}_2(K+Q-Q', K-Q')\Bigr]\biggr\} \,.
\label{eq:2ImPi}
\end{align}

\paragraph*{\underline{Small $q$ expansion.}}
As in the previous section, we expand the integrand in small $q$. 
The electron propagators become:
\begin{align}
G_{K+Q} =&\; 
\frac{1}{\frac{k^2}{2m_*} +\omega - \frac{(\vect{k}+\vect{q})^2}{2m_*}} 
=\frac{1}{\omega -\frac{\vect{k}\cdot\vect{q}}{m_*} -\frac{q^2}{2m_*}}
=\frac{1}{\omega} +\frac{\vect{k}\cdot\vect{q}}{m_*\omega^2} +\dots \,,\\
G_{K-Q'} =&\;
\frac{1}{\frac{k^2}{2m_*} -\omega_{q'} - \frac{(\vect{k}-\vect{q}')^2}{2m_*}} 
= \frac{1}{-\omega +\frac{(\vect{k}+\vect{q}-\vect{q}')^2}{2m_*} - \frac{(\vect{k}-\vect{q}')^2}{2m_*}} 
= -\frac{1}{\omega} -\frac{(\vect{k}-\vect{q}')\cdot\vect{q}}{m_*\omega^2} +\dots\,,
\end{align}
where we have used the energy-conserving delta function to eliminate $\omega_{q'}$ in $G_{K-Q'}$. 
Therefore, at leading order in $q'$,
\begin{align}
&G_{K+Q}\,\widetilde{\mathcal{O}}_1(K, K+Q) + G_{K-Q'} \,\widetilde{\mathcal{O}}_1(K-Q', K+Q-Q') \nonumber\\[4pt]
&=\begin{cases}
G_{K+Q} + G_{K-Q'} \simeq
\dfrac{\vect{q}'\cdot\vect{q}}{m_*\omega^2} & \text{for $\mathcal{O}_1 = \mathbb{1}$}\,,\\
G_{K+Q}\, \dfrac{(2\vect{k}+\vect{q})^2}{8m_e^2} +G_{K-Q'}\,\dfrac{\bigl(2(\vect{k}-\vect{q}')+\vect{q}\bigr)^2}{8m_e^2} \simeq -\dfrac{m_*}{m_e^2}\,\dfrac{\omega-\omega_{q'}}{\omega} & \text{for $\mathcal{O}_1 = \bar v^2$}\,,
\end{cases}
\label{eq:GO}
\end{align}
where an identity operator in spin space is understood, and we have again used energy conservation to simplify the expression in the $\mathcal{O}_1 =\bar v^2$ case. 
Note in particular how the $\mathcal{O}(q^0)$ terms cancel in the case of $\mathcal{O}_1=\mathbb{1}$, such that this LO operator gives a $q$-suppressed contribution. 
The other factor $G_{K+Q}\,\widetilde{\mathcal{O}}_2(K+Q, K) + G_{K-Q'} \,\widetilde{\mathcal{O}}_2(K+Q-Q', K-Q')$ in Eq.~\eqref{eq:2ImPi} is completely analogous, so we obtain, after taking the spin trace (which simply yields a factor of two) and substituting in $y_{q'} = \frac{C_\text{e-ph} q'}{\sqrt{2\omega_{q'}\rho_T^{}}}$, $\omega_{q'}=c_s q'$: 
\begin{align}
\left\{
\begin{matrix}
\text{Im}\,\Pi_{\mathbb{1},\mathbb{1}} \\[4pt]
\text{Im}\,\Pi_{\bar v^2,\mathbb{1}} = \text{Im}\,\Pi_{\mathbb{1},\bar v^2} \\[4pt]
\text{Im}\,\Pi_{\bar v^2,\bar v^2}
\end{matrix}
\right\}
=&\;
-\frac{C_\text{e-ph}^2}{2m_*^2\rho_T^{}c_s}\int\frac{d^3k}{(2\pi)^3} \int\frac{d^3q'}{(2\pi)^3} \, 
2\pi\,\delta\biggl(\omega +\frac{k^2}{2m_*} - \frac{(\vect{k}+\vect{q}-\vect{q}')^2}{2m_*} -\omega_{q'}\biggr)\, \times\nonumber\\
&\hspace{80pt}
f_{\vect{k}}\,\bigl(1-f_{\vect{k}+\vect{q}-\vect{q}'}\bigr)\,
q'\cdot
\left\{
\begin{matrix}
\dfrac{(\vect{q}'\cdot\vect{q})^2}{\omega^4} \\[12pt]
-\dfrac{m_*^2}{m_e^2} \,\dfrac{\vect{q}'\cdot\vect{q}}{\omega^2} \left(1-\dfrac{c_s q'}{\omega}\right) \\[12pt]
\dfrac{m_*^4}{m_e^4} \left(1-\dfrac{c_s q'}{\omega}\right)^2
\end{matrix}
\right\} .
\label{eq:ImPi_result}
\end{align}\\

\paragraph*{\underline{Including the gap.}}
We have presented the calculation of cut diagrams assuming a normal metal for simplicity. 
Accounting for pairing of electrons in the BCS theory introduces a slight modification in the final result in the form of a coherence factor~\cite{PhysRevB.3.305}. 
Concretely, for the imaginary part of two-loop self-energies computed above, this amounts to replacing
\vspace{10pt}
\begin{equation}
2\pi\,\delta\biggl(\omega +\frac{k^2}{2m_*} - \frac{k'^2}{2m_*} -\omega_{q'}\biggr)\,
f_{\vect{k}}\,\bigl(1-f_{\vect{k}'}\bigr) \; \to \;
\frac{\pi}{2} \,\delta\bigl(E_k + E_{k'} +\omega_{q'} - \omega\bigr)\,
\left( 1 - \frac{\epsilon_k \epsilon_{k'} -\Delta^2}{E_k E_{k'}}\right) 
\end{equation}
in Eq.~\eqref{eq:ImPi_result}, where we have abbreviated $\vect{k}+\vect{q}-\vect{q}'\equiv\vect{k}'$ and defined $\epsilon_k \equiv \frac{k^2}{2m_*} -\varepsilon_F^{}$, $E_k\equiv \sqrt{\epsilon_k^2+\Delta^2}$ (and similarly for $\epsilon_{k'}$, $E_{k'}$). 
The energy of the electron-hole pair is therefore constrained to be $E_k+E_{k'}\ge 2\Delta$.\\

\paragraph*{\underline{The $\vect{k}$ integral.}}
We now perform the $\vect{k}$ integral:
\begin{equation}
\mathcal{I} \equiv \int\frac{d^3k}{(2\pi)^3} \,\frac{\pi}{2} \left( 1 - \frac{\epsilon_k \epsilon_{k'} -\Delta^2}{E_k E_{k'}}\right) 
\delta\bigl(E_k + E_{k'} +\omega_{q'} - \omega\bigr) \,.
\end{equation}
The integrand depends only on the magnitude of $k$ and the angle $\theta$ between $\vect{k}$ and $\vect{q}'-\vect{q}$. 
So the azimuthal angle integral simply yields a factor of $2\pi$ and we can use the $\delta$ function to perform the integral over $\cos\theta$. 
The argument of the $\delta$ function has two roots in $\cos\theta$ (corresponding to $\epsilon_{k'} = \pm |\epsilon_{k'}|$), both of which are within the range $[-1, 1]$ in most of viable phase space. 
Noting that $\frac{dE_{k'}}{d\cos\theta} = \frac{\epsilon_{k'}}{E_{k'}} \frac{d\epsilon_{k'}}{d\cos\theta} = -\frac{\epsilon_{k'}}{E_{k'}}\frac{k|\vect{q}'-\vect{q}|}{m_*}$, we have
\begin{equation}
\mathcal{I} = \frac{m_*}{4\pi |\vect{q}'-\vect{q}|}\int_0^\infty dk \,k \,\frac{E_{k'}}{|\epsilon_{k'}|} \left( 1 +\frac{\Delta^2}{E_k E_{k'}}\right) \Theta\bigl( \omega - \omega_{q'} -\Delta -E_k\bigr) \,.
\end{equation}
where $E_{k'} = \omega-\omega_{q'}-E_k$. 
Changing the integration variable from $k$ to $E_k$, we find
\begin{equation}
\mathcal{I} = \frac{m_*^2}{2\pi |\vect{q}'-\vect{q}|} \int_\Delta^{\omega-\omega_{q'}-\Delta} dE \;\frac{EE'+\Delta^2}{|\epsilon\epsilon'|} 
=\frac{m_*^2}{2\pi |\vect{q}'-\vect{q}|} \int_\Delta^{\omega-\omega_{q'}-\Delta} dE \;\frac{EE'+\Delta^2}{\sqrt{(E^2-\Delta^2)(E'^2-\Delta^2)}} \,,
\end{equation}
where a factor of two comes from combining contributions from the two values of $k$ above and below $k_F$ that correspond to the same $E_k$, and we have abbreviated $E_k$, $E_{k'}$, $\epsilon_k$, $\epsilon_{k'}$ to $E, E', \epsilon, \epsilon'$, with $E' = \omega-\omega_{q'}-E$. 
The integral over $E$ can be reduced to elliptic integrals via $E= \frac{1}{2} \bigl[ \omega - \omega_{q'} + t\, (\omega - \omega_{q'} -2\Delta)\bigr]$:
\begin{align}
\mathcal{I} =&\; \frac{m_*^2}{4\pi |\vect{q}'-\vect{q}|} \int_{-1}^1 dt \left[ (\omega-\omega_{q'} +2\Delta) \,\sqrt{\frac{1-\alpha^2t^2}{1-t^2}} -\frac{4\Delta(\omega-\omega_{q'})}{\omega-\omega_{q'}+2\Delta} \,\frac{1}{\sqrt{(1-t^2)(1-\alpha^2t^2)}} \right] \nonumber\\[8pt]
=&\; \frac{m_*^2}{2\pi |\vect{q}'-\vect{q}|} \left[ (\omega-\omega_{q'} +2\Delta)\, E(\alpha) -\frac{4\Delta(\omega-\omega_{q'})}{\omega-\omega_{q'}+2\Delta}\, K(\alpha) \right] \nonumber\\[8pt]
=&\; \frac{m_*^2(\omega-\omega_{q'})}{2\pi |\vect{q}'-\vect{q}|} \left[ (1+\beta)\, E(\alpha) -\frac{2\beta}{1+\beta}\, K(\alpha) \right]
=\frac{m_*^2(\omega-\omega_{q'})}{2\pi |\vect{q}'-\vect{q}|} \,E\Bigl(\sqrt{1-\beta^2}\,\Bigr)
\end{align}
where we have introduced
\begin{equation}
\alpha \equiv \frac{\omega-\omega_{q'}-2\Delta}{\omega-\omega_{q'}+2\Delta} =\frac{1-\beta}{1+\beta} \,,\qquad
\beta \equiv \frac{2\Delta}{\omega-\omega_{q'}} 
\end{equation}
to simplify notation, and
\begin{equation}
K(z) = \int_0^1 dt\, \frac{1}{\sqrt{(1-t^2)(1-z^2t^2)}}\,,\qquad
E(z) = \int_0^1 dt \,\sqrt{\frac{1-z^2t^2}{1-t^2}}
\end{equation}
are the complete elliptic integrals of the first and second kind, respectively. 
In the $\Delta\to 0$ limit, corresponding to a normal conductor, we have $\alpha\to1$, $\beta\to0$, $E(1)=1$, and $\mathcal{I}\to \frac{m_*^2(\omega-\omega_{q'})}{2\pi |\vect{q}'-\vect{q}|}$.

\paragraph*{\underline{The $\vect{q}'$ integral.}}
The remaining integral over the phonon momentum is
\begin{align}
\left\{
\begin{matrix}
\text{Im}\,\Pi_{\mathbb{1},\mathbb{1}} \\[4pt]
\text{Im}\,\Pi_{\bar v^2,\mathbb{1}} = \text{Im}\,\Pi_{\mathbb{1},\bar v^2} \\[4pt]
\text{Im}\,\Pi_{\bar v^2,\bar v^2}
\end{matrix}
\right\}
= &\;
-\frac{C_\text{e-ph}^2\omega}{4\pi \rho_T^{}c_s}\int\frac{d^3q'}{(2\pi)^3} \,E\Biggl(\sqrt{1-\frac{(2\Delta/\omega)^2}{(1-c_s q'/\omega)^2}}\,\Biggr) \times
\nonumber\\
&\hspace{80pt}
\frac{q'}{|\vect{q}'-\vect{q}|} \left(1-\dfrac{c_s q'}{\omega}\right)
\cdot
\left\{
\begin{matrix}
\dfrac{(\vect{q}'\cdot\vect{q})^2}{\omega^4} \\[12pt]
-\dfrac{m_*^2}{m_e^2} \,\dfrac{\vect{q}'\cdot\vect{q}}{\omega^2} \left(1-\dfrac{c_s q'}{\omega}\right) \\[12pt]
\dfrac{m_*^4}{m_e^4} \left(1-\dfrac{c_s q'}{\omega}\right)^2
\end{matrix}
\right\} .
\end{align}
Expanding $\frac{q'}{|\vect{q}'-\vect{q}|}= 1+\frac{\vect{q}'\cdot\vect{q}}{q'^2} +\dots$ and keeping the leading nonvanishing term, we can easily carry out the angular integration. 
Finally, changing the radial integration variable to $x=\frac{c_s q'}{\omega}$, we obtain
\begin{align}
\left\{
\begin{matrix}
\text{Im}\,\Pi_{\mathbb{1},\mathbb{1}} \\[4pt]
\text{Im}\,\Pi_{\bar v^2,\mathbb{1}} = \text{Im}\,\Pi_{\mathbb{1},\bar v^2} \\[4pt]
\text{Im}\,\Pi_{\bar v^2,\bar v^2}
\end{matrix}
\right\}
= &\;
-\frac{C_\text{e-ph}^2\omega^4}{(2\pi)^3\rho_T^{} c_s^4} \int_0^{x_\text{max}} dx \, E\Biggl(\sqrt{1-\frac{(2\Delta/\omega)^2}{(1-x)^2}}\,\Biggr)
\left\{
\begin{matrix}
\dfrac{q^2}{3c_s^2\omega^2}\,x^4(1-x) \\[12pt]
-\dfrac{q^2}{3\omega^2}\, \dfrac{m_*^2}{m_e^2}\, x^2(1-x)^2 \\[12pt]
\dfrac{m_*^4}{m_e^4} \, x^2(1-x)^3
\end{matrix}
\right\} ,
\label{eq:ImPi_final}
\end{align}
where the upper limit
\begin{equation}
x_\text{max} \equiv \min\left(1-\frac{2\Delta}{\omega},\, \frac{\omega_D^{}}{\omega}\right)
\end{equation}
is set by the requirements $\omega-\omega_{q'}\ge 2\Delta$ and $\omega_{q'}=c_s q' \le \omega_D^{}$ (Debye frequency).

When the energy deposition is well above the gap, $\omega\gg 2\Delta$, Eq.~\eqref{eq:ImPi_final} reproduces the normal conductor result:
\begin{align}
\text{Im}\,\Pi_{\mathbb{1},\mathbb{1}} \xrightarrow{\omega\gg 2\Delta} &\; -\frac{C_\text{e-ph}^2}{(2\pi)^3 \rho_T^{}} \,\frac{\omega^2 q^2}{15\,c_s^6} \cdot
\begin{cases}
\dfrac{1}{6} & (\omega\le \omega_D^{}) \,, \\[6pt]
x_D^5 \left( 1 -\dfrac{5}{6}\,x_D^{} \right) & (\omega > \omega_D^{}) \,,
\end{cases} \\[10pt]
\text{Im}\,\Pi_{\bar v^2,\mathbb{1}} = \text{Im}\,\Pi_{\mathbb{1},\bar v^2}  \xrightarrow{\omega\gg 2\Delta} &\; \frac{C_\text{e-ph}^2}{(2\pi)^3 \rho_T^{}}\,\frac{\omega^2q^2}{9\,c_s^4}\, \frac{m_*^2}{m_e^2}\cdot
\begin{cases}
\dfrac{1}{10} & (\omega\le \omega_D^{}) \,, \\[6pt]
x_D^3 \left( 1 -\dfrac{3}{2}\,x_D^{} +\dfrac{3}{5}\,x_D^2 \right) & (\omega > \omega_D^{}) \,,
\end{cases} \\[10pt]
\text{Im}\,\Pi_{\bar v^2, \bar v^2} \xrightarrow{\omega\gg 2\Delta} &\; -\frac{C_\text{e-ph}^2}{(2\pi)^3 \rho_T^{}}\,\frac{\omega^4}{3\,c_s^4}\, \frac{m_*^4}{m_e^4} \cdot
\begin{cases}
\dfrac{1}{20} & (\omega\le \omega_D^{}) \,, \\[6pt]
x_D^3 \left( 1 -\dfrac{9}{4}\,x_D^{} +\dfrac{9}{5}\,x_D^2 -\dfrac{1}{2}\,x_D^3 \right) & (\omega > \omega_D^{}) \,,
\end{cases}
\end{align}
where $x_D^{}\equiv \omega_D^{}/\omega$. 

\paragraph*{\underline{Determination of $C_\text{e-ph}$.}}
We use resistivity measurements~\cite{PhysRevB.36.2920} to determine $C_\text{e-ph}$. 
In Refs.~\cite{PhysRevB.3.305,PhysRevB.36.2920}, a parameter $\lambda_\text{tr}$ is introduced for the electron-phonon coupling, which is defined by
\begin{equation}
\lambda_\text{tr} = 2\int_0^\infty \frac{d\omega'}{\omega'}\, \alpha_\text{tr}^2 \, F(\omega') \,.
\end{equation}
The function $\alpha_\text{tr}^2 \, F(\omega')$ is in turn defined from the conductivity of a normal conductor,
\begin{equation}
\sigma_1(\omega) =\text{Re}\,\sigma(\omega)= \frac{\omega_p^2}{\omega^2}\,\frac{2\pi}{\omega} \int_0^\omega d\omega'\, (\omega-\omega') \,\alpha_\text{tr}^2 \, F(\omega') \,.
\end{equation}
Note that the normalization convention in Ref.~\cite{PhysRevB.3.305} is such that $4\pi\sigma_1$ there equals $\sigma_1$ in our notation. 
From Eqs.~\eqref{eq:dielectric} and \eqref{eq:ImPi_final} (in the limit $\Delta\to 0$) we can readily identify
\begin{equation}
\alpha_\text{tr}^2 \, F(\omega') = 
\begin{cases}
\dfrac{C_\text{e-ph}^2e^2 \omega'^4}{3\,(2\pi)^4\rho_T^{} c_s^6\omega_p^2} & (\omega' \le \omega_D^{})\,,\\[8pt]
0 & (\omega'>\omega_D^{})\,,
\end{cases}
\end{equation}
and therefore
\begin{equation}
\lambda_\text{tr} = \frac{C_\text{e-ph}^2e^2 \omega_D^4}{6\,(2\pi)^4\rho_T^{} c_s^6\omega_p^2} \,.
\end{equation}
For Al, using $\lambda_\text{tr}=0.39$ together with values of the other parameters in Table~\ref{tab:Al-params}, we find $C_\text{e-ph}=56\,$eV.

\section{Absorption in Anisotropic Targets}
\label{app:anisotropic}
Since the benchmark materials considered in this work (Si, Ge and Al-SC) are near-isotropic, in the main text of the paper we worked under the simplifying assumption that the medium is isotropic.
However, it is straightforward to extend the calculation to anisotropic targets. 
In this appendix, we discuss the modifications needed to go beyond the isotropic limit.

First, the in-medium photon self-energy matrix $\Pi_{\lambda\lambda'}$ may have nonzero off-diagonal entries, and its eigenvalues can be found by diagonalization~\cite{Coskuner:2019odd}:
\begin{equation}
\begin{pmatrix}
\Pi_{++} & \Pi_{+-} & \Pi_{+L} \\
\Pi_{-+} & \Pi_{--} & \Pi_{-L} \\
\Pi_{L+} & \Pi_{L-} & \Pi_{LL} 
\end{pmatrix} 
\;\overset{\text{diagonalize}}{\longrightarrow} \;
\begin{pmatrix}
\Pi_1 & 0 & 0 \\
0 & \Pi_2 & 0 \\
0 & 0 & \Pi_3
\end{pmatrix} \,.
\label{eq:rotation}
\end{equation}
A DM state $\phi$ may mix with all three photon polarizations, and Eq.~\eqref{eq:R} generalizes to
\begin{equation}
R = -\frac{\rho_\phi}{\rho_T}\,\frac{1}{\omega^2} \; \text{Im} \left( \Pi_{\phi\phi} +\sum_{i=1,2,3}\frac{\Pi_{\phi i}\Pi_{i\phi}}{m_\phi^2 -\Pi_i^{}}\right)\,,
\label{eq:R_anisotropic}
\end{equation}
where $\Pi_{\phi i}$ is obtained from $\Pi_{\phi A}^\mu$ by first projecting onto $e_{\pm,L}^\mu$ and then rotating into the diagonal basis. 

For vector DM $\phi$, the same rotation in Eq.~\eqref{eq:rotation} diagonalizes also the $\phi\phi$ and $\phi A$ self-energy matrices. 
So each of the three polarizations of $\phi$ mixes only with the one corresponding photon polarization, and we simply replace $\Pi_{T,L}$ in Eq.~\eqref{eq:R} by $\Pi_{1,2,3}$ and average over the three polarizations to obtain the rate in the anisotropic case:
\begin{equation}
R_\text{vector} = -\frac{1}{3}\,\kappa^2\,\frac{\rho_\phi}{\rho_T} \,m_\phi^2 \sum_{i=1,2,3}\text{Im} \left(\frac{1}{m_\phi^2 -\Pi_i^{}}\right)\,. 
\end{equation}

For pseudoscalar DM, still assuming spin-degenerate electronic states, we obtain from Eq.~\eqref{eq:R_ps_1}:
\begin{equation}
R_\text{pseudoscalar} = -g_{aee}^2 \,\frac{\rho_\phi}{\rho_T} \,\frac{1}{4m_e^2\omega^2}\, \frac{1}{e^2} \,\Bigl[\omega^2\,\text{Im}\,(\Pi_{++} +\Pi_{--}) + m_\phi^2\,\text{Im}\,\Pi_{LL}\Bigr] \,,
\end{equation}
which generalizes Eq.~\eqref{eq:R_ps_0}. 
Note that while anisotropy allows for a nonzero mixing between the DM $\phi$ and the photon (via its coupling to the electron's magnetic dipole), its contribution to absorption rate is at $\mathcal{O}(q^2)$ and negligible. 
On the other hand, if the electronic states are not spin-degenerate (\eg\ due to spin-orbit coupling), one would need to explicitly compute additional matrix elements of the spin operator $\vect{\Sigma}$ between the spin part of the wave functions, and the absorption rate cannot be written in terms of components of the photon self-energy matrix. 
Also, mixing between the DM and the photon becomes relevant in this case.

For scalar DM, anisotropy may introduce mixing with all three photon polarizations, and Eq.~\eqref{eq:R_anisotropic} applies. 
The final result, however, is still expected to be dominated by the $\Pi_{\bar v^2,\bar v^2}$ term from $\Pi_{\phi\phi}$, and we therefore have the same formula, Eq.~\eqref{eq:R_s_approx}, as in the isotropic case:
\begin{equation}
R_\text{scalar} \simeq -d_{\phi ee}^2 \, \frac{4\pi m_e^2}{M_\text{Pl}^2}\,\frac{\rho_\phi}{\rho_T}\,\frac{1}{m_\phi^2}\,\text{Im} \,\Pi_{\bar v^2, \bar v^2} \,.
\end{equation}

\bibliographystyle{utphys}
\bibliography{biblio}

\end{document}